\documentclass[fleqn,usenatbib]{mnras}

\usepackage{newtxtext,newtxmath}
\usepackage{pdflscape}
\usepackage{placeins}

\usepackage[T1]{fontenc}

\DeclareRobustCommand{\VAN}[3]{#2}
\let\VANthebibliography\thebibliography
\def\thebibliography{\DeclareRobustCommand{\VAN}[3]{##3}\VANthebibliography}

\def\emerlin{$e$-MERLIN}

\usepackage{graphicx}	
\usepackage{epstopdf}
\usepackage{amsmath}	
\usepackage{multirow}
\usepackage{caption}
\usepackage{longtable}
\usepackage{cprotect}
\usepackage{lscape}
\usepackage{morefloats}
\usepackage{multirow,bigdelim,array,pdflscape,booktabs,psfig}
\usepackage{graphicx} 
\usepackage{subcaption}

\usepackage{etoolbox}
\makeatletter
\patchcmd\@combinedblfloats{\box\@outputbox}{\unvbox\@outputbox}{}{%
}
\makeatother

\title[LeMMINGs VII: The 5 GHz AGN sample]{LeMMINGs VII: 5 GHz, 50 mas \textit{e}-MERLIN observations of a statistically complete sample of nearby AGN}

\author[D.~R.~A. Williams-Baldwin et al.]{D.~R.~A. Williams-Baldwin$^{1}$\thanks{E-mail: david.williams-7@manchester.ac.uk}, 
R.~D. Baldi$^{2}$,
R.~J. Beswick$^{1}$,
I.~M. McHardy$^{3}$,
E.~Carver$^{3}$,
J.~Clifford$^{3}$,
\newauthor 
B.~T.~Dullo$^{4}$,
N.~Kill$^{3}$,
B.~Krishnamoorthi$^{3}$,
I.~M.~Mutie$^{5,1}$,
O.~Woodcock$^{1}$,
M.~K.~Argo$^{6}$,
P.~Boorman$^{7}$, 
\newauthor 
E.~Brinks$^{8}$, 
D.~M.~Fenech$^{9}$,
J.~H.~Knapen$^{10,11}$,
S. Mathur$^{12,13,14}$, 
J.~Moldon$^{15}$,
T.~W.~B.~Muxlow$^{1}$,
\newauthor 
M. Pahari$^{16}$,
N.~H. Wrigley$^{1}$,
A.~Alberdi$^{15}$,
W.~Baan$^{17,18}$,
A.~Beri$^{19,3,20}$,
X. Cheng$^{21}$,
D.~A.~Green$^{22}$,
\newauthor 
J.~Healy$^{23,1}$,
P.~Kharb$^{24}$,
E.~K\"ording$^{25}$,
G.~Lucatelli$^{1}$,
F.~Panessa$^{26}$,
M. Puig-Subir\`a$^{15}$,
C.~Romero-Ca\~nizales$^{27}$,
\newauthor 
D.~J.~Saikia$^{28,29}$,
P.~Saikia$^{30,31}$,
F.~Shankar$^{3}$,
S. Sharma$^{16}$, 
I.~R.~Stevens$^{32}$,
and E.~Varenius$^{1}$
}

\date{Accepted XXX. Received YYY; in original form ZZZ}

\pubyear{2025}

\begin{document}
\label{firstpage}
\pagerange{\pageref{firstpage}--\pageref{lastpage}}
\maketitle

\begin{abstract}

We present 5\,GHz \textit{e}-MERLIN radio images at 50\,mas resolution of the nuclear regions of the Legacy \textit{e}-MERLIN Multi-band Imaging of Nearby Galaxy survey (LeMMINGs): the deepest, statistically complete radio-band survey of the local Universe ($<$120\,Mpc), consisting of 280 galaxies spanning all morphological and nuclear types. We detect nuclear radio emission above a median $5\sigma$ threshold of 0.33\,mJy\,beam$^{-1}$ in 68/280 sources (24\,per\,cent), with core luminosities $10^{35}$--$10^{41.9}$\,erg\,s$^{-1}$. The radio emission is attributed to active galactic nuclei (AGN), circumnuclear star formation, or -- in the case of NGC 3690 -- a tidal disruption event. The brightest radio nuclei, with brightness temperatures $\geq10^{6}$\,K, reside in optically `active' galaxies -- LINERs and Seyferts. The detection rate for `inactive' systems (\ion{H}{II} and absorption-line galaxies), which may host low-luminosity AGN (LLAGN), is 8\,per\,cent. Most detections (78\,per\,cent) are compact ($<$10\,pc), while the remaining 22\,per\,cent show extended, jet-like features (up to 380\,pc). Compared to 1.5\,GHz LeMMINGs data, the 5\,GHz observations' superior resolution and spatial filtering resolve out large-scale structures, isolating genuine nuclear emission. Our results suggests that LLAGN are the primary manifestation of black-hole activity in the local Universe in the form of compact jets and cores, with a preference for early-type hosts. The two LeMMINGs campaigns show that up to 30 per cent of the local galaxy population host a radio-active nucleus, highlighting the necessity of high-resolution, high-sensitivity imaging for uncovering nuclear emission at the lowest luminosities. 

\end{abstract}

\begin{keywords}
galaxies: active – galaxies: jet – galaxies: nuclei – galaxies: star formation – radio continuum: galaxies
\end{keywords}


\newpage
\section{Introduction}
\label{sec:Intro}

A growing body of evidence suggests that a super-massive black hole (SMBH) resides in the centre of nearly all galaxies \citep[e.g., see reviews by][]{Magorrian98,KingPounds}. Active SMBHs -- known as `Active Galactic Nuclei' (AGN) -- play a significant role in galaxy evolution due to kinetic feedback from their powerful jets which regulate galaxy-scale star formation (SF) \citep[e.g.,][]{heckman14,hardcastle20,alexander25}. These jets are often observed at centimeter wavelengths, where the emission is unattenuated by the interstellar medium (ISM) and local absorption of the AGN. Radio emission is a key diagnostic of accretion onto an active SMBH, particularly at low accretion rates where traditional optical or X-ray measurements may be below detection thresholds. 

In the nearby Universe, many of the SMBHs have low accretion rates and are labelled as `low-luminosity AGN' \citep[LLAGN, see e.g., ][for a review]{HoReview}, commonly defined as having $H\alpha$ luminosities $L_{\rm H\alpha} \leq$10$^{40}$ erg s$^{-1}$ \citep{Ho97a} or $L_{\rm X-ray} <$10$^{42}$ erg s$^{-1}$ \citep{Ptak2001}. LLAGN are most likely powered by a radiatively-inefficient accretion flow \citep[`RIAF', see e.g., ][]{NarayanADAF,ho99b,2005MNRAS.363..692D} at significantly sub-Eddington rates \citep[e.g.,][]{Awaki2001PASJ...53..647A,HoReview,KauffmannHeckman2009,2010A&A...516A..87S,2011A&A...527A..22I,2023A&A...670A..22F} resulting in low-level broadband emission \citep{ho09,HoReview,Panessa2007A&A...467..519P}, in contrast to the efficient discs which power traditional (Seyfert-like) AGN and quasars \citep[e.g.,][]{David2011ApJ...728...98D}. 

Despite being unaffected by absorption along the line-of-sight, radio emission in LLAGN can be very weak at GHz frequencies (less than a few mJy) \citep[e.g.,][]{Saikia2018}, contributing minimally to the overall bolometric luminosity and necessitating high-sensitivity observations. As such, they are often `radio-quiet', defined as having a ratio between the 4400{\AA} optical and 6~cm radio emission less than ten \citep{Kellerman1989}. The radio emission in radio-quiet AGN can be attributed to a multitude of processes \citep[see][for a review]{Panessa2019}, including: relativistic or sub-relativistic jets \citep{padovani16,blandford19,Kharb2024}, circum-nuclear SF \citep{condon92}, an outflowing magnetic corona \citep{laor08} or disc winds \citep{zakamska14}. Fortunately, Very Long Baseline Interferometry (VLBI) is capable of resolving the radio emission in the nuclear regions of galaxies from SF and the other processes (e.g., \citealt{falcke00,yan25,saikia25}). Sub-kiloparsec-scale jets with a compact high brightness temperature core ($T_{B} > 10^{7}$K) are a key signature of a LLAGN \citep{morabito22,Cheng2025}. Kiloparsec-scale radio jets have been found in `active' galaxies such as Seyferts and Low-ionisation nuclear emission line regions (LINERs) \citep[e.g.,][]{falcke00,filho02,Gallimore2006,Mundell2009,Panessa2013,Singh2015}, but fewer studies have focused on the `inactive' galaxies like absorption line galaxies (ALGs) and \ion{H}{ii} galaxies which show no evidence for photoionisation from an AGN in their optical spectra. Some `inactive' galaxies may also harbour a weakly accreting active SMBH with sub-kiloparsec-scale jets that could be detected by high-quality sub-arcsecond-resolution radio observations. 

Carrying out an unbiased census of accretion of SMBHs at the lowest accretion rates is of utmost importance as they represent the most common type of AGN in the local Universe \citep{Ptak2001,Nagar2002,filho06,HoReview,Saikia2018}. To perform such a study, we have utilised the best-selected sample of nearby galaxies, the Palomar bright spectroscopic
sample of nearby galaxies \citep[][hereafter referred to as the `Palomar sample']{Filippenko85,Ho95,Ho97a,Ho97b,Ho97c,Ho03,ho09}, as the parent sample of our study. The Palomar sample contains 486 Northern galaxies and is statistically complete to a brightness limit of $B_T < 12.5 \rm  mag$. It includes all galaxy morphologies: early and late-type galaxies (ETGs and LTGs), and all optical nuclear classes: Seyferts, LINERs, ALGs and \ion{H}{ii} galaxies. The high-quality optical spectra from Palomar have been used to probe multi-wavelength correlations of predominantly `active' nuclei to discover radio emission associated with them using various radio interferometers: Karl G Jansky Very Large Array \citep[VLA, $\sim$arcsecond resolution,][]{nagar00,filho00,nagar01,HoUlvestad,Nagar2002,filho02,nagar05,kharb12,balmaverde13,Saikia2018,Chiaraluce2019MNRAS.485.3185C}, MERLIN/\textit{e}-MERLIN \citep[$\sim$100\,mas resolution,][]{filho06,NGC37182007A&A...464..553K} and Very Long Baseline Array (VLBA)/European VLBI Network (EVN) \citep[$\sim$10\,mas resolution,][]{falcke00,ulvestad01b,Nagar2002,anderson04,filho04,nagar05,NGC37182007A&A...464..553K,kharb12,Panessa2013,Cheng2025}. 

\vspace{-2em}
\subsection{The LeMMINGs sample}

The \textbf{L}egacy \textbf{e}-\textbf{M}ERLIN \textbf{M}ulti-band \textbf{I}maging of \textbf{N}earby \textbf{G}alaxies \textbf{S}urvey (LeMMINGs; \citealt{BeswickLemmings}) has produced a statistically complete, sub-arcsecond-resolution and sub-mJy-sensitivity census at 1.5\,GHz of nearby galaxies selected from the Palomar sample, unbiased to other indicators of activity \citep{BaldiLeMMINGs,BaldiLeMMINGs2,BaldiLeMMINGs3}. 
As \emerlin{} is a radio interferometer based in the United Kingdom with six antennas used for this survey (Mark 2 ‘Mk2’, Knockin ‘Kn’, Defford ‘De’, Pickmere  ‘Pi’, Darnhall ‘Da’, and Cambridge ‘Cm’), a declination limit of $\delta >20^\circ$ on the Palomar sample was imposed to avoid highly elliptical synthesized beams. This selection reduces the number of objects to 280 (out of the 486 sources in the Palomar sample) but improves the radio imaging fidelity and crucially does not bias the statistical completeness of the sample. The sample consists of 140 \ion{H}{ii} galaxies, 18 Seyferts, 28 ALG and 94 LINERs, by nuclear AGN type and 15 irregular, 189 spiral, 51 lenticular and 25 elliptical galaxies, by Hubble type.

LeMMINGs comprises two tiers: the `shallow' and the `deep' survey. The `deep' tier focuses on individual interesting Palomar objects \citep[see,][]{MuxlowM82,Williams4151,Westcott2017,Williams6217,Williams4151_2,Dullo5322,RampadarathM51}. Instead, the `shallow' tier uses shorter observations of the 280 Palomar galaxies with multi-wavelength follow-up. The radio observations at 1.5\,GHz at 200\,mas resolution have already been published \citep{BaldiLeMMINGs,BaldiLeMMINGs2,BaldiLeMMINGs3}. In-so-doing, we re-classified each of the galaxies into Seyferts, LINERs, ALGs and \ion{H}{ii} galaxies based on updated Baldwin, Phillips, Terlevich (BPT) diagrams \citep*{BaldwinPhillipsTerlevich81} using the diagnostic tools described in \citet{Kewley06} and \citet{Buttiglione2010} as part of the 1.5\,GHz `shallow' sample release \citep{BaldiLeMMINGs,BaldiLeMMINGs2}. 
We have also compiled multi-wavelength data of matching resolution in the X-ray band with \textit{Chandra} \citep[][Pahari et al., in prep.]{X-rayLeMMINGs} and optical with \textit{Hubble} \citep[][]{LeMMINGsVI,LeMMINGsV,dullo24}. These additional data provide further indicators of nuclear activity, and the multi-wavelength empirical correlations (see \citealt{BaldiLeMMINGs3}) can help to understand the complex nature of the nuclear region in a statistically-relevant sized sample of LLAGN. 

\vspace{-1em}
\subsection{The need for 5 GHz observations of the LeMMINGs sample}

The 1.5\,GHz \emerlin{} images have an average synthesized beam of 200\,mas, compared to the 50\,mas beam size in the 5\,GHz observations presented here, probing linear scales that are four times smaller. At the median distance of the LeMMINGs sample, 20\,Mpc, the angular scale is $\sim$100~pc/arcsec, corresponding to linear sizes of 5\,pc at the resolution of our 5\,GHz \emerlin{} beam-size. The higher resolution used in the 5\,GHz data should result in a relatively `clean' sample of compact nuclear radio emission attributed solely to AGN activity, unaffected by other processes (diffuse SF, large-scale jets/outflows), which are expected to be resolved out. 

A second observing band also provides a two-point spectral-index measurement. The radio spectrum $\alpha$, is defined as S$_{\rm \nu} \propto \nu^{\alpha}$. For a flat radio spectrum, $\alpha \sim$0, a steep radio spectrum $\alpha <$0, and an `inverted' radio spectrum $\alpha >$0. Radio emission from the compact radio nuclei associated with LLAGN often has a flat or inverted spectrum \citep[e.g.,][]{Nagar2002}, whereas the radio spectrum of jets is expected to be steeper due synchrotron ageing of electrons \citep[see e.g.,][]{HoReview}. Thus, a compact flat-spectrum radio source with high brightness temperature provides conclusive proof of an (LL)AGN. 

In this manuscript we present the 5\,GHz data release of the nuclear regions of all 280 LeMMINGs galaxies in order to compare with the 1.5\,GHz data. This work represents the 5\,GHz data release of the LeMMINGs sample and as such we focus solely on the radio image quality of the nuclear regions of these galaxies. Future work will include optical and X-ray information to perform a multi-wavelength study of the nuclear regions, as well as presenting a full wide-field catalogue of all sources in the 1.5\,GHz and 5\,GHz images to study off-nuclear radio emission. 
This paper is structured as follows: in Section~\ref{sec:section2}, we describe the 5\,GHz \emerlin{} observations, reduction and imaging procedures. In Section~\ref{sec:results} we show the results from the survey. In Section~\ref{sec:discussion}, we discuss these results in comparison with the 1.5\,GHz data and finally, in Section~\ref{sec:conclusions} we summarise our results and present our conclusions.

\vspace{-1.5em}
\section{Observations and Data Reduction}
\label{sec:section2}
\subsection{Optical positions of the LeMMINGs survey galaxies}
\label{sec:Gaia}

Before discussing the 5\,GHz \textit{e}-MERLIN radio data, we first provide updated optical positions for all the sources in the sample. Previously in the 1.5\,GHz sample, we obtained the best optical positions of all LeMMINGs sources based on the NASA Extragalactic Database (NED)\footnote{\url{https://ned.ipac.caltech.edu/}} or the Simbad online database\footnote{\url{https://simbad.u-strasbg.fr/simbad/sim-fbasic}} \citep{simbad}. Since the data release of the 1.5\,GHz LeMMINGs data \citep{BaldiLeMMINGs,BaldiLeMMINGs2}, optical positions of many galaxies have been significantly improved by the \textit{Gaia} satellite \citep{gaia2016A&A...595A...1G,gaia2017A&C....21...22S,gaia2018A&A...616A...1G}, resulting in high-precision positions of many of the LeMMINGs sample. 

The positions of high-resolution radio VLBI data have been shown to have little disagreement with \textit{Gaia} positions. For example, \citet*{Petrov2017} found that only 6 per cent of sources had a VLBI-\textit{Gaia} offset above a 99 per cent confidence level (the median position offset in their sample was 2.2 mas), although they showed that genuine offsets can exist. For the cases where the VLBI and \textit{Gaia} positions are significantly offset from one another, the disagreement can be attributed to the detection of a radio jet that is brighter than the core \citep[see e.g.,][]{popkov25}. In nearby galaxies, the agreement between \textit{Gaia} and catalogued NED/Simbad source positions is within the quoted uncertainties for more than 80 per cent of targets, but when galaxies are considered on their own, removing the stars, quasars, white dwarfs and binary stars from the analysis, this figure falls slightly \citep{Hales2024}. The reason for the difference between the  NED/Simbad literature and \textit{Gaia} source positions is found to be due to a mismatch in classifications, for example galaxies classified by \textit{Gaia} as stars, or stars classified by \textit{Gaia} as quasars \citep{Hales2024}. Furthermore, foreground objects can also cause a misidentification of a nearby galaxy \citep{Barmby2023MNRAS.518.3746B}. Therefore, one must not rely on the \textit{Gaia} positions, but also compare with literature positions of nearby galaxies to pin-point a nuclear region in which to search for radio emission.

When querying the \textit{Gaia} database\footnote{\url{https://gaia.ari.uni-heidelberg.de/singlesource.html}.} for the LeMMINGs galaxies, 68/280 (24 per cent) of the positions are identical to the best positions provided by NED or Simbad. 205/280 (73 per cent) agree to within less than an arcsecond. A recent VLBI study of a sub-sample of LeMMINGs galaxies \citep{Cheng2025} has shown that detected low-luminosity AGN with high brightness temperatures are found within 1-arcsec ($<$100\,pc on the sky) of the \textit{Gaia} optical position. Similarly, Karl G Jansky Very Large Array (VLA) observations of the Palomar sample identified radio cores at a resolution of $\sim 1$ to $2$ arcseconds \citep{filho00,Nagar2002,nagar05}. Of the 75 galaxies that do not agree to within 1-arcsecond of the \textit{Gaia} position, there is no obvious pattern in which sources tend to be mis-matched when considering their optical nuclear class or galaxy type. In terms of optical classifications, 45/140 \ion{H}{ii} galaxies (32 per cent), 5/18 Seyferts (28 per cent), 8/28 ALGs (29 per cent) and 17/94 LINERs (18 per cent) have \textit{Gaia} optical positions that are discrepant with those in Simbad by more than 1 arcsecond. With regards to the morphological galaxy types, the positions of 4 irregular galaxies (15, 27 per cent), 56 spirals (189, 30 per cent), 4 lenticulars (51, 8 per cent) and 11 ellipticals (25, 44 per cent) are discrepant by more than 1 arcsecond.

In the 1.5\,GHz data release, we classified sources as identified if significant radio emission associated with the `radio core' of an AGN was found within 1.5-arcsec of the galaxy centre from the best optical positions available from the literature \citep{BaldiLeMMINGs,BaldiLeMMINGs2}. This larger radius was used because of the worse astrometry and positional accuracy of order $\sim 1$ to $2$ arcseconds of the Two Mass All Sky Survey data\footnote{\url{https://old.ipac.caltech.edu/2mass/releases/second/doc/sec6 7f.html}} which was available for the source positions. As the \textit{e}-MERLIN astrometry is tied to the International Celestial Reference Frame and correct to 10\,mas accuracy, the positional accuracy is dominated by the optical/IR data. Thus the offset of 1.5-arcsec was used as a conservative maximum offset in which to search for radio cores to pinpoint the active SMBH. 

Using 1.5-arcsec as the radius for searching for radio emission, in total 224/280 galaxies have \textit{Gaia} optical positions that agree with those found on Simbad/NED and therefore we use the \textit{Gaia} position to centre our images (see Section~\ref{sec:selfcal}). For the other 56 sources, we use the position obtained from Simbad/NED to centre our images. However, noting that significant offsets between \textit{Gaia} and Simbad/NED can exist, we also checked for radio emission at the position of the \textit{Gaia} optical centre for all sources with significant discrepancy between the \textit{Gaia} and Simbad/NED positions. In total, we detect radio emission within 1.5-arcsec of the \textit{Gaia} optical centre in 60 sources, with the remaining eight detected sources being within 1.5-arcsec of the Simbad/NED position. We further discuss the impact of this position choice in Section~\ref{sec:identified} and ~\ref{sec:Gaiaoffsetsdiscussion}.

\subsection{5 GHz LeMMINGs observations}

Observations of the LeMMINGs sample centred at 5.07\,GHz (C band) took place between December 2018 and September 2019. Additional observations of sources that were missed or provided poor data quality in the original observing runs were obtained in a further observation in November 2020. In all observations, six of the seven \emerlin{} antennas were available for observations, with the Lovell not taking part. In some observations, various technical issues with individual antennas may have prevented them from observing for parts of the observing runs, but we ensure that the data are of similar minimum quality. We give a subjective indication of data quality for each dataset in Table~\ref{tab:obslog}, and an indication of the image quality for each source in Table~\ref{tab:basictable}. 

The 280 objects were grouped into 28 observing blocks by their separation on the sky to reduce telescope slewing time. We used the standard \textit{e}-MERLIN `C-band': a single band of 512\,MHz in width between 4.82 and 5.33\,GHz. Four spectral windows across the band were used, each with 512 channels of 0.25\,MHz resolution, prior to averaging. Each target field was observed for $\sim$50 minutes in total using a `snap-shot' imaging mode, whereby each field was visited once per hour for eight hours. This fills up a larger portion of the \textit{uv}-plane and provides a more circular synthesized beam and thus increase imaging fidelity. Each object had a nearby unresolved and bright phase calibrator which was observed for 2 minutes either side of 6-minute scans on the target. The flux calibrator (3C286) and band pass calibrator (OQ208) fields were observed at the beginning or end of each observing block. In some cases, LeMMINGs blocks were split in two for scheduling efficiency or due to problems with the array which reduced observing time on previous imaging runs. These runs were combined into one observation to obtain a sensitivity level commensurate with the other datasets. This strategy was similar to that used for the 1.5\,GHz sample.

\vspace{-1em}
\subsection{Data reduction and calibration}

The 5\,GHz \emerlin{} data were calibrated with version 5.6.2 of the Common Astronomy Software Applications \textsc{CASA} package \citep{CASA} using version 1.1.19 of the \emerlin{} \textsc{CASA} Pipeline \citep[eMCP,][]{eMCP2}. The same pipeline was used in the second 1.5\,GHz release of LeMMINGs data \citep{BaldiLeMMINGs2}.  The eMCP\footnote{Full descriptions of the calibration steps performed by the eMCP can be found on the online github pages: \url{https://github.com/e-MERLIN/eMERLIN_CASA_pipeline}}. runs in two sections: pre-processing and calibration. The pre-processing step loads the data into \textsc{CASA} measurement set format, performs preliminary data flagging, including observatory correlator flags, \textit{a priori} flagging on known areas of radio frequency interference (RFI) in the data. Automated flagging of RFI with \textsc{AOFlagger} \citep{AOFlagger} is disabled for the 5\,GHz data as the RFI environment is generally much cleaner at 5\,GHz. Post flagging, the data are averaged to a 4-second integration time, and by a factor of 4 in frequency, so that there are 128 channels per spectral window. The flag version state is then saved in preparation for the calibration procedures. 

The calibration section employs standard radio calibration techniques, starting with the inclusion of a manual flag file that can be added by the user, then performing an initial band pass and delay calibration. A flux model of the flux calibrator 3C286, which is slightly resolved on \emerlin{} baselines, is used to perform flux calibration. The radio spectral index across the band for all calibrator fields it then computed. Using this new band pass information, the previous bandpass, phase, and amplitude calibration steps are repeated to correct for this spectral index across the band. These calibrations are applied to the data and a final round of automated flagging using \texttt{tfcrop} in \textsc{CASA} is performed on the target fields to remove any low-level RFI. Calibration plots are produced and stored in the eMCP weblog for further inspection. Basic, preliminary images of the target and phase calibrator fields are produced. Phase calibrator and target fields are split into their own measurement sets, though averaged further down, so that quick images can be made by the user. 

\begin{table}

    \centering
    \caption{Observing log of the 5\,GHz \textit{e}-MERLIN data presented in this work. 
    The data quality flag is as follows: `++'=good, `+'=adequate. A `good' dataset is defined as one where all 6 antennas are available for the observation throughout most of the run, whereas `adequate' datasets are those where one or more antennas drop out for more than half an observing run. The datasets denoted with a `\textdagger' are those which are two shorter observing runs that have been combined and calibrated together. }
    \begin{tabular}{c c c c}
    \hline
        LeM Block & Obs. Date & Antennas & Data Quality \\
        \hline
        LEMTEST & 2018-12-06 & Mk2KnDePiDaCm & ++ \\
        LEM02 & 2019-08-21 & Mk2KnDePiDaCm & ++ \\
        LEM03 & 2018-12-10 & Mk2KnDePiDaCm & ++ \\
        LEM04 & 2018-12-09 & Mk2KnDePiDaCm & ++ \\
        LEM05 & 2019-08-13 \textdagger & Mk2KnDePiDaCm & ++ \\
        LEM06 & 2019-08-23 & Mk2KnDePiDaCm & ++ \\
        LEM07 & 2019-07-09 & KnDePiDaCm & + \\
        LEM08 & 2018-12-03 & Mk2KnDePiDaCm & ++ \\
        LEM09 & 2019-08-30 & Mk2KnDePiDaCm & + \\
        LEM10 & 2019-08-12 & Mk2KnDePiDaCm & ++ \\
        LEM11 & 2019-08-25 & Mk2KnDePiDaCm & ++ \\
        LEM12 & 2019-08-24 & Mk2KnDePiDaCm & ++ \\
        LEM13 & 2019-08-15 & Mk2KnDePiDaCm & ++ \\
        LEM14 & 2019-07-18 & Mk2KnDePiDaCm & ++ \\
        LEM15 & 2019-07-19 & Mk2KnDePiCm & + \\
        LEM16 & 2018-12-08 & Mk2KnDePiDaCm & ++ \\
        LEM17 & 2019-08-09 \textdagger & Mk2KnDePiDaCm & + \\
        LEM18 & 2019-07-19 & Mk2KnDePiCm & + \\
        LEM19 & 2019-08-04 & Mk2KnDePiCm & + \\
        LEM20 & 2019-08-02 & Mk2KnDePiDaCm & ++ \\
        LEM21 & 2019-08-22 & Mk2KnDePiDaCm & ++ \\
        LEM22 & 2019-08-03 & Mk2KnDePiDaCm & ++ \\
        LEM23 & 2019-08-15 & Mk2KnDePiDaCm & ++ \\
        LEM24 & 2019-08-24 \textdagger & Mk2KnDePiDaCm & ++ \\
        LEM25 & 2019-08-25 & Mk2KnDePiDaCm & ++ \\
        LEM26 & 2018-12-05 & Mk2KnDePiDaCm & ++ \\
        LEM27 & 2019-08-04 & Mk2KnDePiCm & + \\
        LEM28 & 2019-08-22 & Mk2KnDePiDaCm & ++ \\
        LEMmisc & 2020-10-16 & Mk2KnDePiDaCm & ++ \\
        \hline
    \end{tabular}
    \label{tab:obslog}
\end{table}

\vspace{-1em}
\subsection{Imaging and self calibration}
\label{sec:selfcal}

\begin{figure}
    \centering
    \includegraphics[width=0.95\columnwidth]{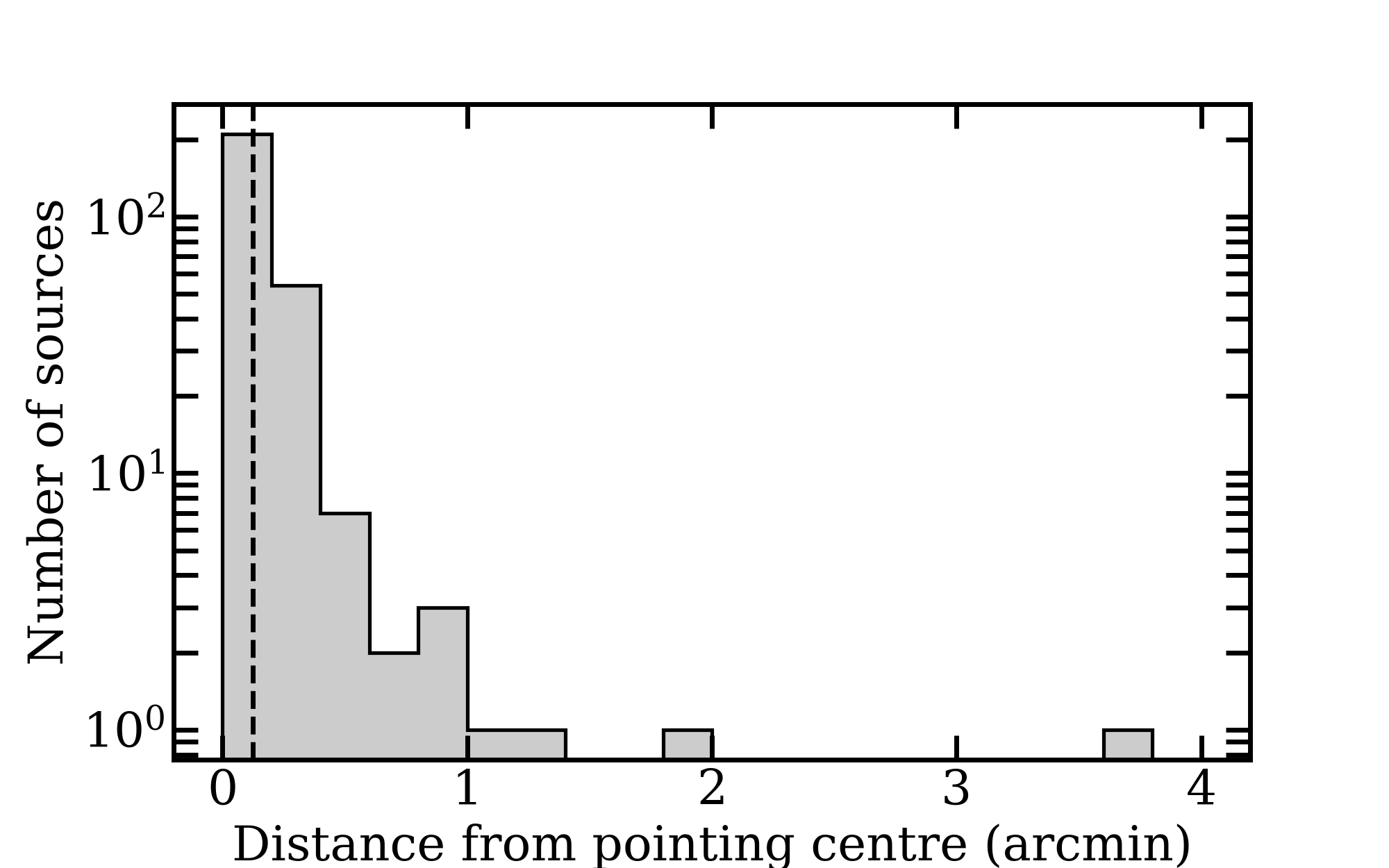}
    \caption{The distribution of optical position offsets relative to the \textit{e}-MERLIN observation pointing positions for all 280 sources in the sample. Four sources are further than 1-arminute away from the central pointing, with only NGC~5055 ($>$3-arcmin away) being re-processed in a specific way to ameliorate the problems of smearing (see Section~\ref{sec:selfcal}). The vertical dashed line is the median offset of the sample which is 7.5\,arcsec.}
    \label{fig:distfrompointingcenter}
\end{figure}

We constructed a semi-automated imaging pipeline to remove any additional RFI in the calibrated data and perform self calibration of the data, using bright sources in the primary beam. This additional work to self-calibrate using nearby bright sources is desirable for the following reasons: i) many of the sources in this survey are likely to be faint ($\lesssim$1~mJy), but a bright ($\gtrsim$1~mJy) in-beam source, when present, could provide a good candidate for self-calibration to improve the overall data quality of the object we are interested in; ii) a catalogue of sources detected within the larger primary beam will be the subject of a future LeMMINGs paper; iii) the legacy value of having created high-resolution wide-field radio images of 280 nearby galaxies for future users; iv) many of the galaxy nuclei are not at the beam pointing centre due to previous inaccurate catalogues used for the original positions of the sources. On this final point, we show a histogram of the observation pointing positions relative to the optical centres of the LeMMINGs galaxies in Fig.~\ref{fig:distfrompointingcenter}. In most cases, the optical positions agree with the \textit{e}-MERLIN pointing centres to within 1 arcmin, which is inside the $\approx$7-arcmin (FWHM) at 5\,GHz primary beam for \textit{e}-MERLIN without the Lovell telescope. However, for four of sources, this is not the case: NGC~4169 (1.06 arcmin), NGC~3600 (1.20 arcmin),  and NGC~5395 (1.82 arcmin) and NGC~5055 (3.74 arcmin). For NGC~5055, as the source position is significantly offset from the pointing centre, we re-processed the data for this source on its own, phase rotating the data at full frequency and time resolution to the correct position of NGC~5055, to help with the self-calibration of the source. To minimise the effects of the non-central pointing, we also performed a primary beam correction.

We used a combination of the imaging routines in \textsc{wsclean} \citep{offringa-wsclean-2014,offringa-wsclean-2017} and \textsc{CASA} to create full-resolution self-calibrated and primary beam corrected images of all 280 sources. To perform this task, we first image the entire seven arcmin \emerlin{} primary beam with \textsc{wsclean} at low resolution (one pixel per synthesized beam) with uniform weighting in order to find the bright compact sources $>$1\,mJy in the image that may be suitable for self-calibration. Using the rudimentary \textsc{CASA} source finding algorithm\footnote{See the \textsc{findsources} task here: \url{https://casadocs.readthedocs.io/en/stable/api/tt/casatools.image.html}}, we produced an outlier field file of these bright sources. This file was used in \textsc{CASA} to perform standard imaging and self-calibration procedures on smaller cut-out images to reduce computing resources. For NGC~3034, we used a list of known compact radio sources above 1~mJy from \citet{fenech08} to perform the self-calibration. First, we performed five rounds of  phase self-calibration, reducing the solution interval down to 180~s. In seven cases (NGC~3031, NGC~3516, NGC~315, NGC~4138, NGC~4151, NGC~1167 and NGC~1275), a source in the beam (or the target itself) was bright enough to perform additional rounds of self-calibration down to 24~s in phase, and apply an `amplitude+phase' self-calibration correction. In other cases where the source(s) in the field were not brighter than 1 mJy (NGC~783, NGC~5055, NGC~4274, NGC~3884, NGC~4369, NGC~3190, NGC~4102, NGC~3838, NGC~3945, NGC~4648, NGC~2985, IC~520, IC~356, NGC~2500, NGC~2964, NGC~2683, NGC~5033), we performed three phase-only self-calibration loops to a solution interval of 360~s. If we found no sources in the widefield low-resolution image, we did not perform self-calibration. In total, we were able to self-calibrate 70/280 fields, improving the sensitivity levels in these images. 

After self-calibration, we imaged each field with \textsc{WSClean}, and we produced a final Nyquist sampled image (0.008 arcsec pixel size) of the central 4-arcmin $\times$ 4-arcmin with Briggs weighting \citep[][]{Briggs} with robust=0.5. We used a bespoke \textsc{wsclean} container that includes the \textit{e}-MERLIN primary beam models to produce the final images to correct for the primary beam. We also used a slight Gaussian taper (\texttt{-taper-gaussian 0.04 arcsec}) with \textsc{wsclean} to help circularise the beam without degrading the noise level significantly. For NGC~4151 and NGC~315 we were unable to use this parameter without making the synthesized beam significantly larger ($>$100 mas). In order to compare the radio emission on scales similar to that in the 1.5\,GHz data release, we made further tapered images with a Gaussian taper of 0.1 arcseconds, which resulted in an average beam size of 140\,mas. 

\begin{figure}
    \centering
    \includegraphics[width=0.95\columnwidth]{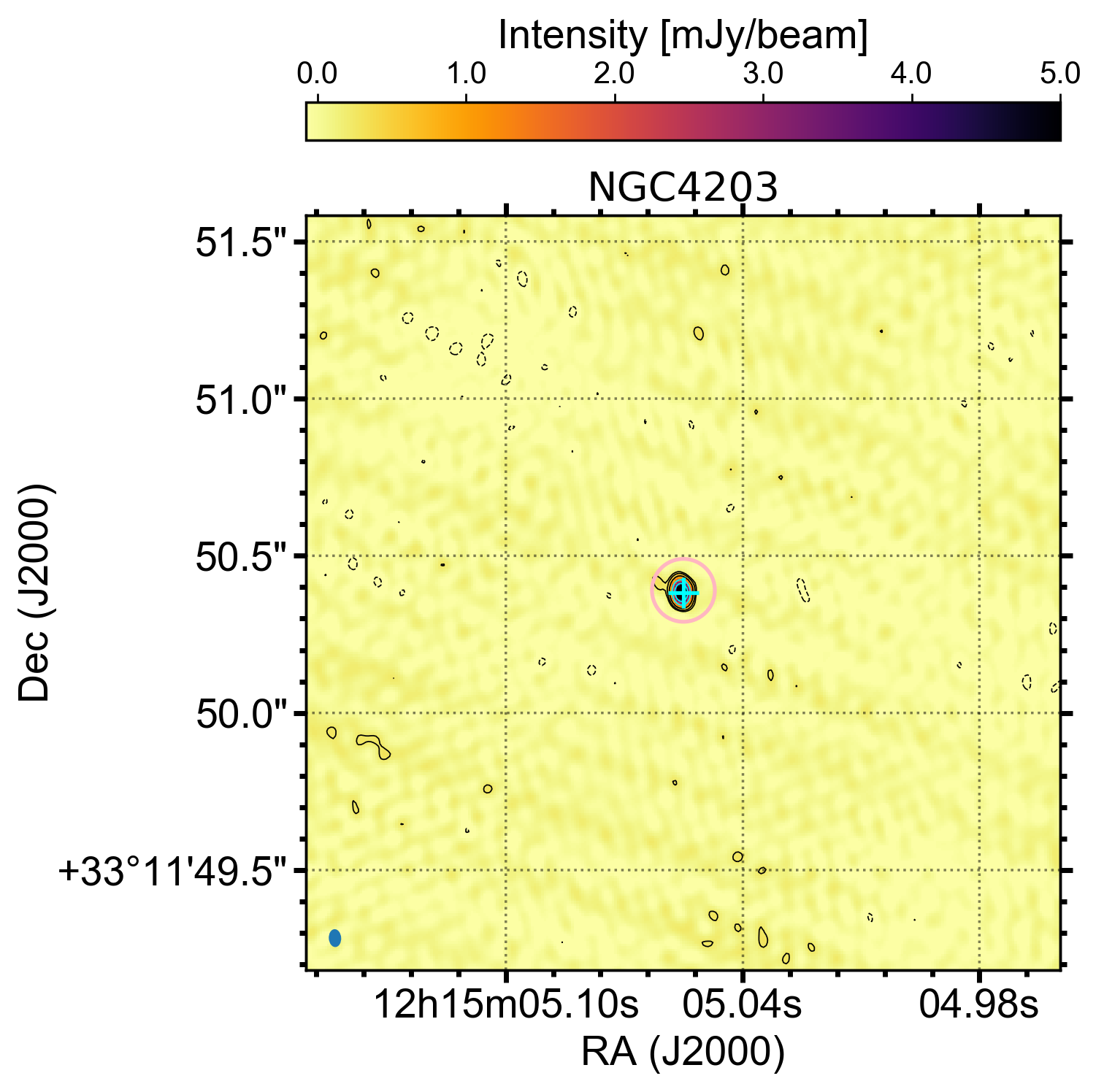}
    \caption{Central 2.5$\times$2.5 arcsec$^{2}$ of the 5\,GHz radio image of NGC\,4203 as an example of the LeMMINGs 5\,GHz data, made using \textsc{APLpy} \citep[][]{APLpy}. See 
    online supplementary material for all the images. The false colour background image ranges from 0 to 5\,mJy beam$^{-1}$, with the colour bar shown at the top. The contour levels are set at the image r.m.s. (in this case 77~$\upmu$Jy beam$^{-1}$) multiplied by $-$3, 3, 5 (black contours) and 10, 50, 100, 250 (light blue contours). Negative contours are dashed. The light-blue plus represents the Gaia DR3 optical position, where available (Section~\ref{sec:Gaia}). 
    The pink circle denotes the average 1.5\,GHz synthesized beam positioned at the 1.5\,GHz `core' position. The synthesized beam is shown in the bottom left corner of the image. 
    }
    \label{fig:NGC4203example}
\end{figure}

\begin{figure}
    \centering
    \includegraphics[width=0.95\columnwidth]{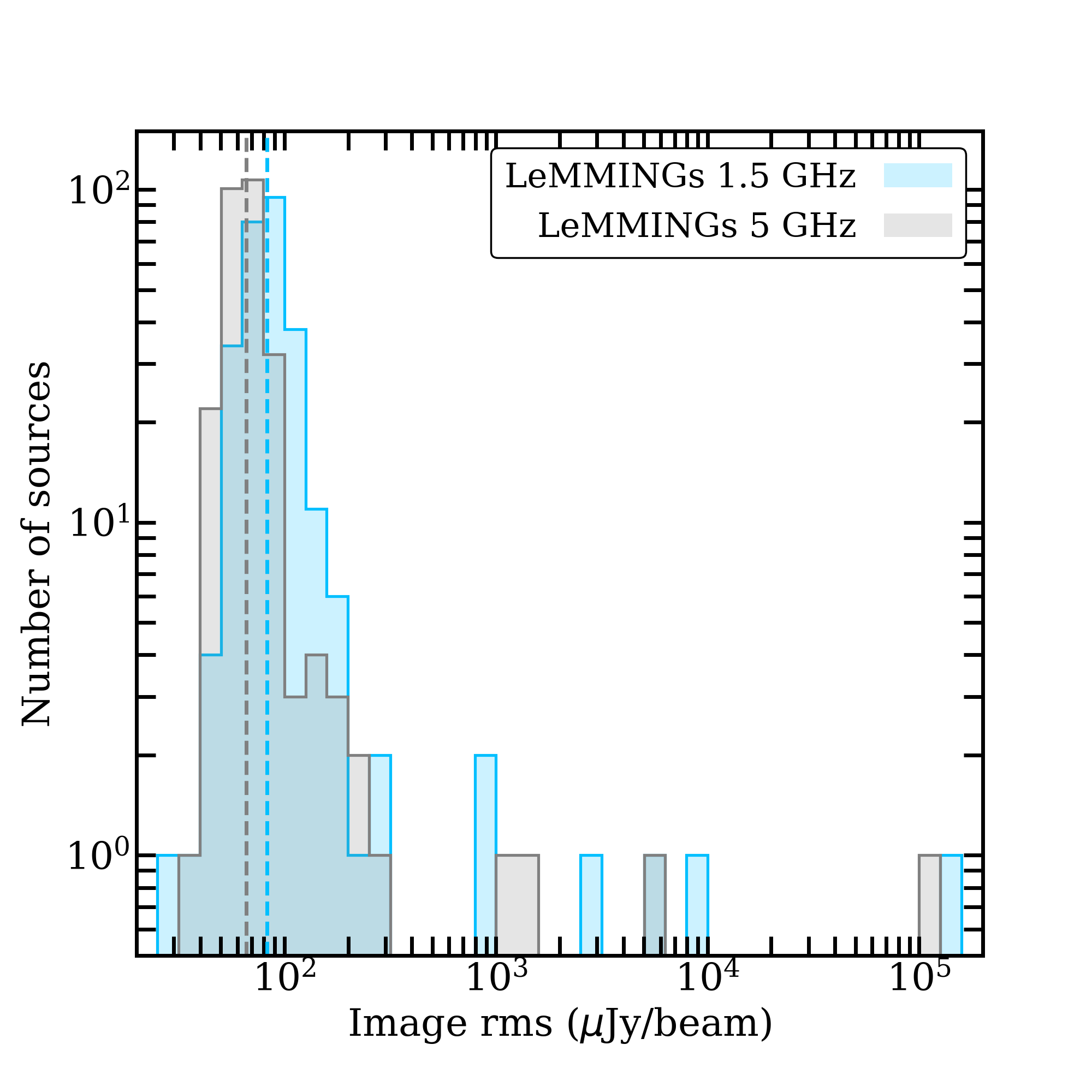}
    \caption{Histograms of the image r.m.s. distribution of the noise in the LeMMINGs C-Band (5\,GHz, grey) and L-Band (1.5\,GHz, blue) samples. The median of the distributions are marked with dashed lines, corresponding to 66~$\upmu$Jy beam$^{-1}$ at 5\,GHz and 83~$\upmu$Jy beam$^{-1}$ at 1.5\,GHz.}
    \label{fig:rmsdistribution}
\end{figure}

We also made smaller `cut-out' images of the central part of each galaxy for both the full-resolution and tapered images centred at the best optical position (see Section~\ref{sec:Gaia}). We present these images in the online supplementary material, but an example full-resolution image of NGC\,4203 is shown in Fig.~\ref{fig:NGC4203example}. The figures were made using \textsc{APLpy} \citep[][]{APLpy} and show a 2.5$\times$2.5 arcsec$^{2}$ section centred on the nucleus. We also made larger images and found extended large-scale jets in several cases (see Section~\ref{sec:largescalejets}), and can be found in the online supplementary material. Overall the data and image quality in the majority of fields are objectively good. We deemed that 236, 38 and 8 have  respectively `good',  `adequate'  and  `poor' quality images, respectively. As a consequence, we obtain a median r.m.s. noise level of our 5\,GHz images of 66~$\upmu$Jy beam$^{-1}$ at 5\,GHz. This compares favourably to the median noise of 83~$\upmu$Jy beam$^{-1}$ for the 1.5\,GHz sample (see Fig.~\ref{fig:rmsdistribution}) \citep{BaldiLeMMINGs2}. The 5\,GHz noise level is an order of magnitude improvement over similar high-resolution radio surveys of the Palomar survey \citep[e.g.,][]{Nagar2002}.

\vspace{-1em}
\subsection{Parameter extraction}
\label{sec:paramsextract}

We used \textsc{CASA} to extract source parameters from the final images of all sources. We estimated the noise using the task \texttt{imstat} in an off-source region near to the optical centre of the galaxy. We used the task \texttt{imfit} to extract flux densities, sizes and positions of detected radio components. In sources where multiple radio components were separated within 1-beam width of each other (NGC~1167, NGC~2273, NGC~2655, NGC~3348, NGC~4151 and NGC~4589), we fitted the brightest component first to find the position and flux density, which we here-after designate as the `core'. We then fitted the other components with a fixed synthesized beam size to extract their parameters, by also fixing the flux and position of the main bright component. This enabled us to derive reliable intensity values for fainter radio sources. 
Where a source is not detected in the tapered image but is detected in the full-resolution image (or vice versa), we give a 5$\sigma$ upper limit for that component. If a source is confused with another source in the tapered image, we do not give a low-resolution flux for that component. Instead, it is included as part of the brighter component of the two. From the fitted flux densities of OQ208 in all observing runs, we estimated a 5 per cent flux calibration uncertainty in our data, which we include in quadrature to our flux uncertainties derived from the source fitting. The flux densities for all detected sources are shown in Table~\ref{tab:fluxbasic} (A machine readable version of the full table can be found in the online supplementary material.).

\vspace{-1em}
\section{Results}
\label{sec:results}

\subsection{Radio images and source parameters}
\label{sec:identified}

We aim to find the compact radio sources associated with the centres of nearby galaxies where SMBHs could reside, and often circum-nuclear SF is found as well \citep{dullo24}. Therefore, we have searched the optical position of the centre of each galaxy for significant radio emission above a 5$\sigma$ level of the local r.m.s. sensitivity of each source. In contrast, for the 1.5\,GHz data we preferred a less conservative approach by providing a 3$\sigma$ detection threshold for tentative identifications \citep{BaldiLeMMINGs2}. In this work we choose the higher threshold to provide more certain detections. For example, other \textit{e}-MERLIN legacy surveys such as e-MERGE used thresholds such as 4.8$\sigma$ to reduce the contamination by false detections \citep{e-MERGE}. This sets the median detection threshold for the 5\,GHz LeMMINGs dataset is 0.33~mJy beam$^{-1}$.

In the 1.5\,GHz LeMMINGs sample \citep{BaldiLeMMINGs,BaldiLeMMINGs2}, we calculated that one random source would be detected above a 3$\sigma$ level (0.25~mJy) within a 3.7 and 2.6 arcsec circular error radii, when comparing with the deep \textit{XMM-Newton}/\textit{ROSAT} data \citep{seymour04} and the \textit{e}-MERLIN SuperCLASS survey \citep{Battye2020}, respectively. Analogously, based on the source count distribution at 5\,GHz from the ATLAS 5.5 GHz survey of the extended Chandra Deep Field South \citep[][3.5 arcsec resolution over an area 0.34 deg$^2$]{2015MNRAS.454..952H}, we find that when observing 280 galaxies, statistically at most one unrelated radio source falls within a radius of $\sim$4.3 arcsec of the optical source. Furthermore, the probability of a detection at the optical position in our radio images is given by the error function \citep*[e.g., see Section 3.1 of][]{Kording2005}. We calculate that the likelihood of a 5$\sigma$ detection within 1.5 arcsec of the optical position is 0.05 per cent. These analyses all validate our choice to use a 1.5 arcsec search radius for finding nuclear radio emission at 1.5 and 5 GHz associated with the optical centre of the galaxy.

\clearpage

\onecolumn

\begin{landscape}
{\renewcommand{\arraystretch}{1.2}%
\begin{center}
\begin{longtable}{ c c c c c c c c c c c c c c c c c}

\caption{First five lines of the table of basic properties of the 5\,GHz radio sample presented in this paper. The full table can be found in the online supplementary material. 
The columns correspond to: (1) galaxy name; (2) source name used in scheduling and for retrieving images; (3) J2000.0 Right Ascension; (4) J2000.0 Declination; (5) Distance in Mpc; (6) LeMMINGs Observing Block. Note that any block denoted with an asterisk was re-observed as part of the miscellaneous LeMMINGs block; (7) Image quality, with quality rating: `++'=good, `+'=adequate; (8) r.m.s. noise level in $\upmu$Jy beam$^{-1}$; (9) Synthesized beam size in arcseconds; (10) Synthesized beam size position angle in degrees; (11) Detection at 1.5\,GHz: Detected = `D', Undetected = `U'; (12) Morphology of the 5\,GHz data. `A' = core, `B' = single/one-sided jet/extension, `C' = double-sided jets, `E' = multiple/complex set of components; (13) optical nuclear class, from the classification of \citep[][]{BaldiLeMMINGs,BaldiLeMMINGs2}. Note that `LINER' has been shortened to `LIN.' and `Seyfert' to `Sey.'; (14) Galaxy morphological type from the classification of \citep[][]{BaldiLeMMINGs,BaldiLeMMINGs2}: Spiral (`Sp'), Lenticular (`Len.'), Elliptical (`El.') and Irregular (`Irr.'); (15) log 5\,GHz radio luminosity. Sources with a preceding `$<$' denotes a 5$\sigma$ upper limit. } 
\label{tab:basictable} \\

\hline
Galaxy & Source & Right Asc. & Dec. & Dist.    & LEM   & Image.  & r.m.s.     & Beam & Beam     & 1.5\,GHz  & morph. & AGN  & Gal. & log(L$_{\rm 5\,GHz}$)\\
Name.  & Name. & $\alpha$ J2000.0  &  $\delta$ J2000.0 & (Mpc) & Block   & Quality & $\upmu$Jy/b & Size ($\arcsec$)  & P.A. (deg)     & Det. &  & Type & Type & erg s$^{-1}$\\
(1)    &  (2)  & (3)  & (4)      & (5)   & (6)   &  (7)    & (8)  & (9)      & (10)     & (11) & (12)   & (13) & (14) & (15) \\
\hline
\endfirsthead

\multicolumn{16}{c}%
{\tablename\ \thetable\ -- \textit{Continued from previous page}} \\
\hline
Galaxy & Source & Right Asc. & Dec. & Dist.    & LEM   & Image.  & r.m.s.     & Beam & Beam     & 1.5\,GHz  & morph. & AGN  & Gal. & Log(L$_{\rm 5\,GHz}$)\\
Name.  & Name. & $\alpha$ J2000.0  &  $\delta$ J2000.0 & (Mpc) & Block   & Quality & $\upmu$Jy/b & Size ($\arcsec$)  & P.A. (deg)     & Det. &  & Type & Type & erg s$^{-1}$\\
(1)    &  (2)  & (3)  & (4)      & (5)   & (6)   &  (7)    & (8)  & (9)      & (10)     & (11) & (12)   & (13) & (14) & (15) \\
\hline
\endhead
\hline \multicolumn{16}{r}{\textit{Continued on next page}} \\
\endfoot
\hline
\endlastfoot

NGC~7817 &   0003+2045 &    00$^{\rm h}$03$^{\rm m}$59$\fs$198 & +20$^{\circ}$45$\arcmin$10$\farcs$92 &       31.5 &           04 &          ++ &      73 &           0.08$\times$0.05 &  26.59 &           U &                             $-$ &      HII &       Sp.  & $<$36.34\\
IC~10 &   0020+5917 &    00$^{\rm h}$20$^{\rm m}$24$\fs$993 & +59$^{\circ}$17$\arcmin$17$\farcs$91 &        1.3 &           04 &          ++ &      73 &           0.06$\times$0.05 &  34.71 &           U &                             $-$ &      HII &    Irr.  & $<$33.57\\
NGC~147 &   0033+4830 &    00$^{\rm h}$33$^{\rm m}$12$\fs$171 & +48$^{\circ}$30$\arcmin$32$\farcs$17 &        0.7 &           04 &          ++ &      76 &           0.06$\times$0.06 &  33.06 &           U &                             $-$ &      ALG &    Irr.  & $<$33.05\\
NGC~185 &   0038+4820 &    00$^{\rm h}$38$^{\rm m}$57$\fs$940 & +48$^{\circ}$20$\arcmin$15$\farcs$04 &        0.7 &           04 &          ++ &      72 &           0.06$\times$0.05 &  25.59 &           U &                             $-$ &    LIN. &    Irr.  & $<$33.02\\
NGC~205 &   0040+4141 &   00$^{\rm h}$40$^{\rm m}$22$\fs$055 & +41$^{\circ}$41$\arcmin$07$\farcs$50 &        0.7 &           04 &          ++ &      71 &           0.06$\times$0.05 &  43.42 &           U &                             $-$ &      ALG &    Irr.  & $<$33.02\\

\end{longtable}
\end{center}
}

\vspace{-3em}
\begin{small}
{\renewcommand{\arraystretch}{1.2}%
\begin{center}

\begin{longtable}{ c | c c c c c c c c | c c c c | c }

\caption{First four lines of the detected sources table. The full table can be found in the online supplementary material. 
Column description: (1) galaxy name; (2) radio component: core or additional component denoted `add.', with `W' or `E' standing for West or East, respectively; (3) deconvolved FWHM dimensions (major × minor axes arcsec$^{2}$, $\Theta_{\rm M} \times \Theta_{\rm m}$) of the fitted component, determined from a 2D Gaussian fit on the full-resolution radio image. If the value has an asterisk `*' next to it, it was fitted using \texttt{imfit} while forcing the fit size to the synthesized beam (see Section~\ref{sec:paramsextract}; (4) PA of the deconvolved component, PA$_{\rm d}$, from the full-resolution radio image (degree); (5) r.m.s. of the radio image close to the specific component from the full-resolution radio image ($\upmu$Jy beam$^{-1}$); (6) and (7) radio position (J2000.0); (8) peak intensity in mJy beam$^{-1}$, F$_{\rm peak}$, from the full-resolution radio image; (9) integrated flux density, F$_{\rm total}$, in mJy, from the full-resolution radio image; (10) deconvolved FWHM dimensions (major $\times$ minor axes arcsec$^{2}$, $\Theta_{\rm M} \times \Theta_{\rm m}$) of the fitted component, determined from a 2D Gaussian fit on the low-resolution radio image. Where a source is resolved in the full-resolution image and unresolved in the tapered image, no information is given for the fainter source, but the total flux is included in the brighter component; (11) PA of the deconvolved component, PA$_{\rm d}$, from the low-resolution radio image (degrees); (12) r.m.s. of the radio image close to the specific component from the low-resolution radio image ($\upmu$Jy beam$^{-1}$); (13) peak intensity in mJy beam$^{-1}$, F$_{\rm peak}$, from the low-resolution radio image. The total flux density of the radio source associated with the galaxy is given in mJy beneath the horizontal line, measured from the low-resolution image.; (14) radio morphology (A, B, C, E) and size in arcsec and pc.} \label{tab:fluxbasic} \\

\hline
 & \multicolumn{8}{c}{Full-resolution} & \multicolumn{4}{c}{Tapered-resolution} & \multicolumn{1}{c}{}
\\
\hline
Galaxy & com. & $\Theta_{\rm M} \times \Theta_{\rm m}$ & PA$_{\rm d}$ & r.m.s.    & Right Asc.   & Dec.  & F$_{\rm peak}$     & F$_{\rm total}$  & $\Theta_{\rm M} \times \Theta_{\rm m}$ & PA$_{\rm d}$ & r.m.s. &   F$_{\rm peak}$ & morph/size\\
Name.  &      & arcsec$^2$  &  deg. & $\upmu$Jy~beam$^{-1}$ & $\alpha$ J2000.0   & $\delta$ J2000.0 & mJy~beam$^{-1}$ & mJy & arcsec$^2$  &  deg. & $\upmu$Jy~beam$^{-1}$ &    mJy~beam$^{-1}$ & \\
(1)    &  (2)  & (3)  & (4)      & (5)   & (6)   &  (7)    & (8)  & (9)      & (10)     & (11) & (12)   & (13) & (14) \\
\hline
\endfirsthead

\multicolumn{14}{c}%
{\tablename\ \thetable\ -- \textit{Continued from previous page}} \\
\hline
 & \multicolumn{8}{c}{Full-resolution} & \multicolumn{4}{c}{Tapered-resolution} & \multicolumn{1}{c}{}
\\
\hline
Galaxy & comp & $\Theta_{\rm M} \times \Theta_{\rm m}$ & PA$_{\rm d}$ & r.m.s.    & Right Asc.   & Dec.  & F$_{\rm peak}$     & F$_{\rm total}$  & $\varTheta_{\rm M} \times \varTheta_{\rm m}$ & PA$_{\rm d}$ & r.m.s. &   F$_{\rm peak}$ & morph/size\\
Name.  &      & arcsec$^2$  &  deg. & $\upmu$Jy/b & $\alpha$ J2000.0   & $\delta$ J2000.0 & mJy/b & mJy & arcsec$^2$  &  deg. & $\upmu$Jy/b &    mJy/b & \\
(1)    &  (2)  & (3)  & (4)      & (5)   & (6)   &  (7)    & (8)  & (9)      & (10)     & (11) & (12)   & (13) & (14)\\
\hline
\endhead
\hline \multicolumn{13}{r}{\textit{Continued on next page}} \\
\endfoot
\hline
\endlastfoot

NGC~266 &        core &  $<$0.07$\times<$0.06 &   22.22 &      58 &   00$^{\rm h}$49$^{\rm m}$47$\fs$814 & +32$^{\circ}$16$\arcmin$39$\farcs$75 &   1.15$\pm$0.07 &   1.15$\pm$0.08 & $<$0.24$\times<$0.11 &      $-$41.93 &      77 &        1.00$\pm$0.06 &                              core (A) \\
\cline{13-13}
 &  & & & & & & & & & & & 1.00$\pm$0.07  \\
\hline
NGC~315 &        core &  $<$0.08$\times<$0.07 &  $-$10.76 &    6000 &   00$^{\rm h}$57$^{\rm m}$48$\fs$882 & +30$^{\circ}$21$\arcmin$08$\farcs$83 &   629$\pm$32 &   629$\pm$32 &     $<$0.45$\times<$0.22 &      $-$29.87 &    1685 &      578$\pm$29 &                          core (A) \\
\cline{13-13}
 &  & & & & & & & & & & & 578$\pm$29  \\
\hline
NGC~410 &        core &  $<$0.07$\times<$0.05 &   33.25 &      70 &   01$^{\rm h}$10$^{\rm m}$58$\fs$900 & +33$^{\circ}$09$\arcmin$07$\farcs$02 &   0.63$\pm$0.04 &   0.63$\pm$0.05 &     $<$0.31$\times<$0.11 &      $-$37.64 &      96 &    0.91$\pm$0.09 &                          core (A) \\
\cline{13-13}
 &  & & & & & & & & & & & 0.98$\pm$0.17  \\
\hline
NGC~507 &        core &  $<$0.07$\times<$0.05 &   32.03 &      82 &   01$^{\rm h}$23$^{\rm m}$39$\fs$928 & +33$^{\circ}$15$\arcmin$21$\farcs$79 &   1.24$\pm$0.08 &    1.26$\pm$0.1 &      $<$0.3$\times<$0.11 &      $-$37.28 &      99 &    1.55$\pm$0.09 &                          core (A) \\
\cline{13-13}
 &  & & & & & & & & & & & 1.6$\pm$0.1  \\
\hline

\end{longtable}
\end{center}
}
\end{small}

\end{landscape}

\twocolumn

\vspace{-1em}
We detected significant radio emission above a 5$\sigma$ local r.m.s. and within 1.5 arcsec of the optical galaxy centre for 68/280 targets. For these sources we also derive the peak and total flux densities, positions and deconvolved beam sizes in the full-resolution and \textit{uv}-tapered images, listed in the online supplementary material. We list the contour levels for each of the images in the online supplementary material. 
We use the proximity to the 1.5\,GHz radio core (and to the optical centre) together with the radio properties (peak intensity and compactness) to determine, among multiple detected components, which one is the nuclear radio core associated with the SMBH. For the remaining 212 sources, no significant nuclear emission was detected in either the full-resolution or tapered images. We list image r.m.s. noise values for these non-detections in Table~\ref{tab:basictable}.

In the 1.5\,GHz LeMMINGs data, the detected radio sources were split into different morphological classes based on the appearance of additional nuclear radio emission. Here, we classified the radio morphologies of the 68 detected (identified) sources into five different regimes: compact `core' sources (A), one-sided jets (B), triple sources (C), double-lobed (D) and complex sources (E). We refer the reader to \citet{BaldiLeMMINGs} for a description of these classes and examples. We analogously classify our sources and find that no sources are consistent with the `double-lobed' morphology. In total, we detect 52 `core' morphology sources (A), 10 with one-sided jets (B), 4 triple sources (C) and 1 complex source (E) and list the break down by optical nuclear type in Table~\ref{tab:fraction2}. For all additional radio components found in the detected sources, we list their parameters in Table~\ref{tab:fluxbasic}.

The 5\,GHz data, being of higher angular resolution, are sensitive to smaller-scale structures than the 1.5\,GHz data \citep{BaldiLeMMINGs,BaldiLeMMINGs2} or the previous VLA observations of the Palomar galaxy sample \citep[e.g.,][]{filho00,HoUlvestad,Nagar2002,nagar05}. In addition, jet emission not associated with the core, SF regions, and supernova (SN) remnants all tend to have optically-thin spectra, which reduces the chance of detection at 5\,GHz (see also Section~\ref{sec:Gaiaoffsetsdiscussion} for a discussion). For this reason, we would expect to detect fewer extended structures in our 5\,GHz images when compared to the 1.5\,GHz images. In the 1.5\,GHz data, a total of 40/106 (38 per cent) detected sources were found to have radio morphologies B/C/D/E, i.e., elongation not directly associated with the unresolved `core' (Table~\ref{tab:fraction2}). Instead, in the 5\,GHz sample presented here, we only find 15/68 (22 per cent) sources which show the presence of additional structure. NGG~2782 is the only source with a ring-like, complex (E) radio morphology, and a possible SF nature. The remaining 14 B/C-classified detected sources 
show extended structures and additional components and are referred henceforth as the `jetted' galaxies\footnote{The 14
sources that show evidence of jetted structures (B/C class) are: NGC~1167 (B), NGC~2273 (B), NGC~2639 (B), NGC~2683 (B), NGC~2655 (C), NGC~2841 (B), NGC~3348 (B), NGC~3504 (C), NGC~3516 (B), NGC~4036 (B), NGC~4151 (C), NGC~4589 (C), NGC~4736 (B), and NGC~6703 (B).}. But the origin of this extended emission still needs more investigation. These have radio sizes ranging from 5 to 380 parsecs (Table~\ref{tab:fluxbasic}) and are discussed individually in Section~\ref{sec:jetteddiscussion}.  The detected unresolved (`A') sources, and most of the additional
components in the ‘jetted’ galaxies have approximately the same size
as the e-MERLIN synthesized beam, of the order of $\sim$10 parsec, on average.

\begin{table}
	\centering
	\caption{Radio morphological classification breakdown of the LeMMINGs sample as a function of optical nuclear type and number of detections. The sample is divided into morphological radio (core/core-jet, one-sided jet, triple, double-lobed source, and complex source) and spectroscopic optical classes (LINER, ALG, Seyfert, \ion{H}{II} galaxies) based on their radio detection, core-identification or non-detection. The numbers in parentheses represent the 1.5\,GHz sample values \citep{BaldiLeMMINGs2}. }
	\label{tab:fraction2}
	\begin{tabular}{l|cccc|c} 
		\hline
                                   & \multicolumn{4}{c|}{AGN Type} \\
\hline
Radio class          & LINER & ALG & Seyfert  &  H{\sc ii}  &  Total  \\
		\hline

A (core)     &   35 (37)  &   2 (3)  &    6 (6)  &   10 (18)  & 53 (64) \\
B (extension)     &   7 (2)  &   1 (0)  &    2 (1)  &   0 (2)  & 10 (5) \\
C (triple)     &   2 (13)  &   0 (2)  &    1 (3)  &   1 (4)  & 4 (22) \\
D (2-lobe)    &   0 (3)  &   0 (0)  &    0 (1)  &   0 (0)  &  0 (4) \\
E (complex)     &   0 (1)  &   0 (0)  &    0 (1)  &   1 (9)  & 1 (11) \\
\cmidrule(r){1-6}
Total$_{\rm (identified)}$ &   44 (56)  &   3 (5)  &    9 (12)  &   12 (33)  & 68 (106) \\
\midrule
unidentified    &   0 (2)  &   0 (2)  &    0 (1)  &   0 (14)  & 0 (19) \\    
\midrule
Total$_{\rm (detected)}$ &   44 (58)  &   3 (7)  &    9 (13)  &   12 (47)  & 68 (125) \\
undetected    &   50 (36)  &   25 (21)  &    9 (5)  &   128 (93)  & 212 (155) \\
	\hline\hline
Total           &   94 (94)  &   28 (28)  &    18 (18)  &   140 (140)  & 280 (280) \\	
\hline
\end{tabular}

\end{table}

\subsubsection{Identified sources}

All nuclei are detected in the full-resolution images except for NGC~5194, where we detect the core in the tapered image. All the images can be found in the online supplementary material. 
The median core peak intensity of the detected sources is 1.5\,mJy beam$^{-1}$, but the sample ranges from 0.31\,mJy beam$^{-1}$ (NGC~3735) to 25.3\,Jy beam$^{-1}$ (NGC~1275). 
For the sources labelled as compact `core' sources (A), the radio peak intensity and the flux density agree to within a factor of $\leq$2, with the exceptions being NGC~5194, NGC~6946 and NGC~3729, where the core is slightly resolved in all cases, but no clear jet components are detected. For all the `jetted' sources, the total flux and peak intensities agree to within a factor of 3, with the exception being NGC~4151, which we discuss further in Section~\ref{sec:jetteddiscussion}.

In the 1.5\,GHz sample we classified nineteen sources detected within the innermost 0.73 arcmin $\times$ 0.73 arcmin region but not coinciding with the central optical position as `unidentified' \citep{BaldiLeMMINGs2}. We prefer to discuss the off-nuclear sources in a further work and present only the nuclear-detected sources in this work. However, we note that one source, NGC~3690, is part of an interacting system also called Arp~299. The detected source at 5 GHz corresponds to the component Arp~299-B (see \citealt{BaldiLeMMINGs2} for a discussion), which is the location of a tidal disruption event \citep{Arp299TDE}. Due to this nature, it was classified as `unidentified' in the 1.5\,GHz sample but we include it in this work to provide a complete census of nuclear detections. In this work, it is detected at a peak intensity of 2.79$\pm$0.24~mJy beam$^{-1}$ on 2019 August 03.

\vspace{-1em}
\subsubsection{Undetected sources}

Of the 280 galaxies, 212 were undetected (76 per cent), and no additional sources were found within the central 1.5$\times$1.5 arcsec$^{2}$ region that we searched for radio emission. Information about the image r.m.s. noise levels of these undetected sources can be found in Table~\ref{tab:basictable}. We note that the lack of a detection does not necessarily preclude a weakly emitting AGN. The 5$\sigma$ detection threshold used in this work (0.33~mJy beam$^{-1}$) is higher than the 3$\sigma$ detection threshold (0.25~mJy beam$^{-1}$) used in the 1.5\,GHz sample, thus requiring objects close to the 1.5\,GHz detection threshold to have radio spectral index ($F_{\nu} \sim \nu^{\alpha}$) $\alpha \gtrsim$ 0.23. For compact radio emitting LLAGN or SF, the radio spectrum is expected from flat ($\alpha \approx$ 0) to steep (optically-thin emission, $\alpha \approx -$0.7, e.g., \citealt{falcke00}), depending on whether the emission originates from a synchrotron self-absorbed compact radio `core' or the associated jets, or extended SF. While a detailed comparison between the 1.5\,GHz and 5\,GHz data will be discussed in a forthcoming work, we briefly discuss this effect more in Section~\ref{sec:specindex} and ~\ref{sec:detectionLLAGNdiscussion}.

\vspace{-1em}
\subsection{Radio brightness}

\begin{figure}
\centering
   \begin{subfigure}[b]{\columnwidth}
   \includegraphics[width=0.95\columnwidth]{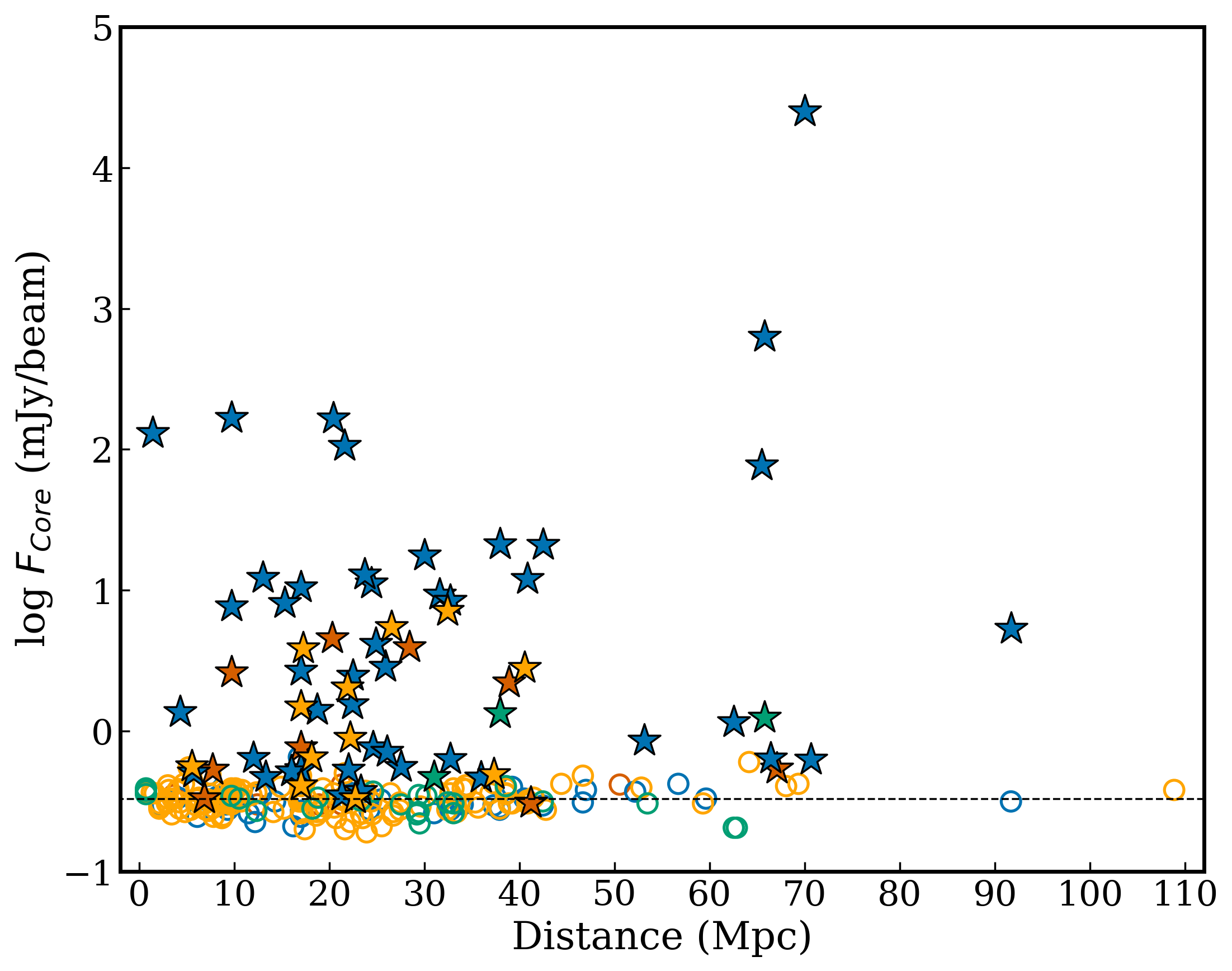}
   \caption{}
   \label{fig:fluxversusdistanceplota} 
\end{subfigure}

\begin{subfigure}[b]{\columnwidth}
   \includegraphics[width=0.95\columnwidth]{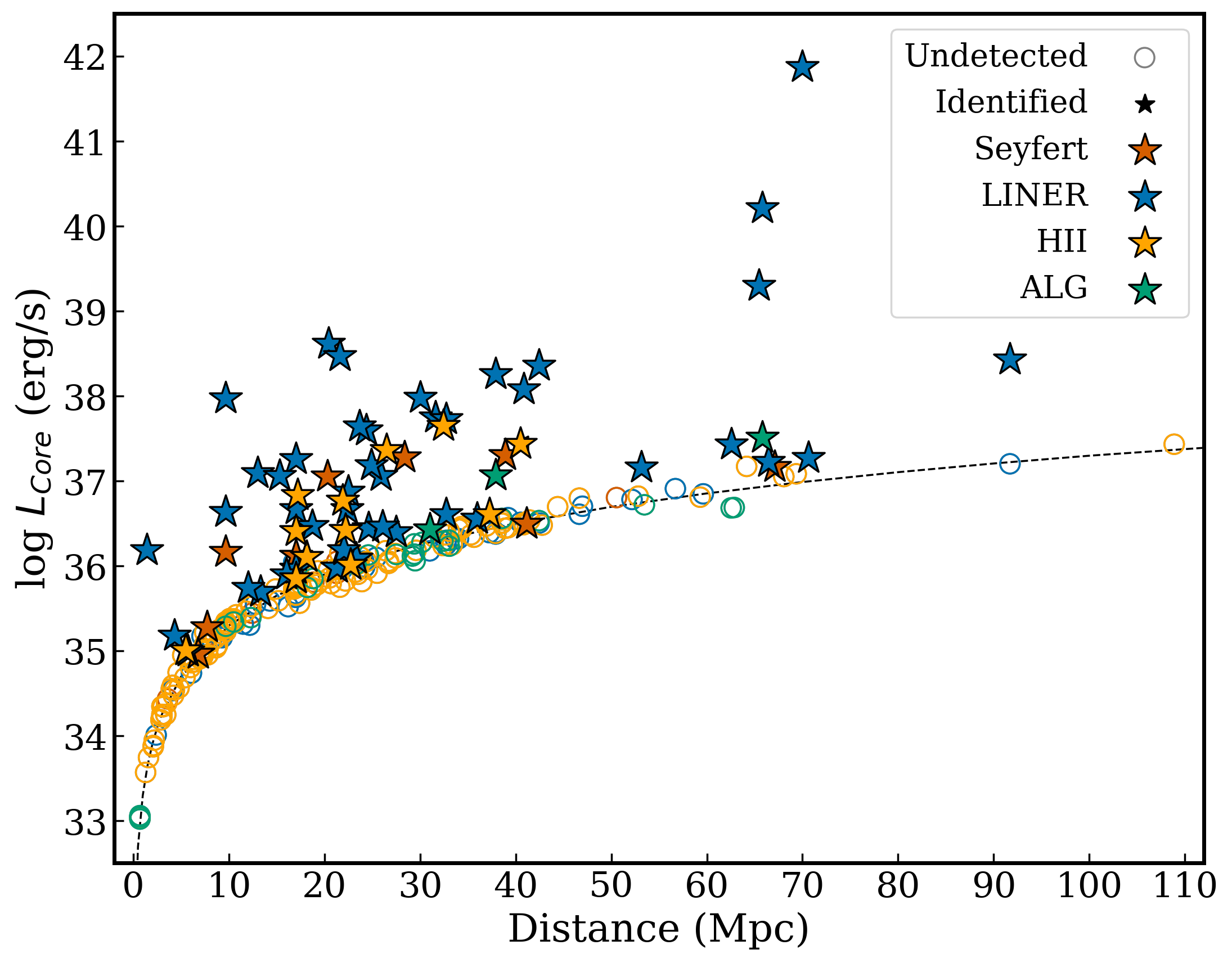}
   \caption{}
   \label{fig:fluxversusdistanceplotb}
\end{subfigure}
\caption{Radio core peak intensity (\textit{panel a}) and luminosity (\textit{panel b}), using the peak of the brightest component plotted as a function of distance (Mpc) for the full sample. The dashed lines represent the 5$\sigma$ flux density limit of the 5\,GHz survey (0.33 mJy beam$^{\rm -1}$). }
    \label{fig:fluxversusdistanceplot}

\end{figure}

We detect compact radio emission from the nuclei of 68 nearby galaxies as part of this 5\,GHz data release of the LeMMINGs survey with peak intensities $\gtrsim$ 0.33\,mJy beam$^{-1}$ (see Fig.~\ref{fig:fluxversusdistanceplot}). The observations achieve a 50\,mas resolution, consistent with a linear scale of 5~pc at the 20\,Mpc median distance of the sample. We detect radio sources up to a distance of 90 Mpc of all nuclear AGN types. Only two sources are more distant and not detected. Six sources have peak intensities in excess of 100\,mJy beam$^{-1}$ and a further nine sources are between 10$-$100\,mJy beam$^{-1}$. These fifteen `bright' sources are dominated by LINERs: all sources above 7.5\,mJy are LINERs (see Fig.~\ref{fig:fluxversusdistanceplot}, \textit{top} plot). Twenty-eight of the detected sources in our sample have peak intensities up to 1\,mJy beam$^{-1}$, showing that over a third of these sources would not have been detected in other similar observations of the Palomar sample with poorer sensitivity \citep[e.g., the 5$\sigma$ threshold of the VLA data in][is 1\,mJy beam$^{-1}$]{Nagar2002}. 

The core radio luminosities\footnote{To convert the luminosities presented here in erg s$^{-1}$ to monochromatic luminosities (W~Hz$^{-1}$), a value 16.7 must be subtracted from the logarithm of the luminosities.} span from 10$^{35}$ to 10$^{41.9}$ erg s$^{-1}$ ($\sim10^{18.3}$ to 10$^{25.2}$~W Hz$^{-1}$), with a median value of 7.1$\times$10$^{36}$ erg s$^{-1}$, an order of magnitude higher than the 1.5\,GHz median luminosity of 5.8$\times$10$^{35}$ erg s$^{-1}$.
This difference between the two surveys can be explained by the higher detection threshold used in the 5\,GHz dataset presented here\footnote{Including only the 75 sources above a 5$\sigma$ detection threshold in the 1.5\,GHz dataset, their median 1.5-GHz luminosity would be 6.3$\times$10$^{36}$ erg s$^{-1}$, similar to the median 5-GHz radio luminosity of the sample.}, and the far lower number of detected radio sources in the 5\,GHz data. The observed radio luminosity range is about an order of magnitude lower than that obtained with previous VLA \citep{Nagar2002} and MERLIN \citep{filho06} studies of the Palomar sample, respectively. The larger sample, the higher resolution and higher sensitivity of our 1.5 and 5-GHz LeMMINGs survey promotes this study as one of the deepest statistically complete campaigns of nuclear radio activity performed in the local Universe.

Similar to the 1.5\,GHz sample, we calculated the brightness temperature ($T_{\rm B}$) of each of the detected sources. Using the 5$\sigma$ detection threshold of 0.33\,mJy beam$^{-1}$ and the typical \textit{e}-MERLIN synthesized beam of 50\,mas, we calculate $T_{\rm B}$ = 6.5$\times$10$^{3}$K, far below the threshold of $\sim$10$^5$K commonly used to discriminate non-thermal AGN from free-free (Bremsstrahlung) emission \citep{condon92}, though care must be taken as radio SN and young SN remnants can also reach up to $\sim$10$^7$K \citep[e.g.,][]{Brunthaler2009A&A...497..103B}. 
For many compact components the intrinsic sizes are not reliably resolved at 50\,mas. As such, the deconvolved sizes reported by \texttt{imfit} should be interpreted as model-fit estimates (often serving as upper limits) rather than robust, fully resolved dimensions. 
If a value of 15\,mas -- a value reported by CASA for some of the most compact sources -- is used instead,  to achieve a brightness temperature of $>$10$^{5}$K, a flux density of 0.5\,mJy beam$^{-1}$ is required. 
There are 55 sources that meet this threshold. Furthermore, if one uses a more conservative thermal/non-thermal  $T_{\rm B}$ separation $>$10$^{6}$K as we did in the 1.5\,GHz sample, then a flux density of 5\,mJy is required. There are 22 sources above this limit, similar to the 21 sources detected above this limit at 1.5\,GHz. All are LINER-type nuclei except for the \ion{H}{II} galaxies NGC~3504 and NGC~3665. Both of these sources have been previously studied: NGC~3504 appears extended in VLBI images \citep{DellerMiddelberg} and may have a weak LINER nucleus \citep{2000ApJ...530..688A}, whereas NGC~3665 is also classified as a Fanaroff--Riley type-1 radio galaxy because of the 3~kpc-scale jets that are seen with the VLA \citep{parma86}.

\begin{figure}
\centering
   \begin{subfigure}[b]{\columnwidth}
   \includegraphics[width=0.95\columnwidth]{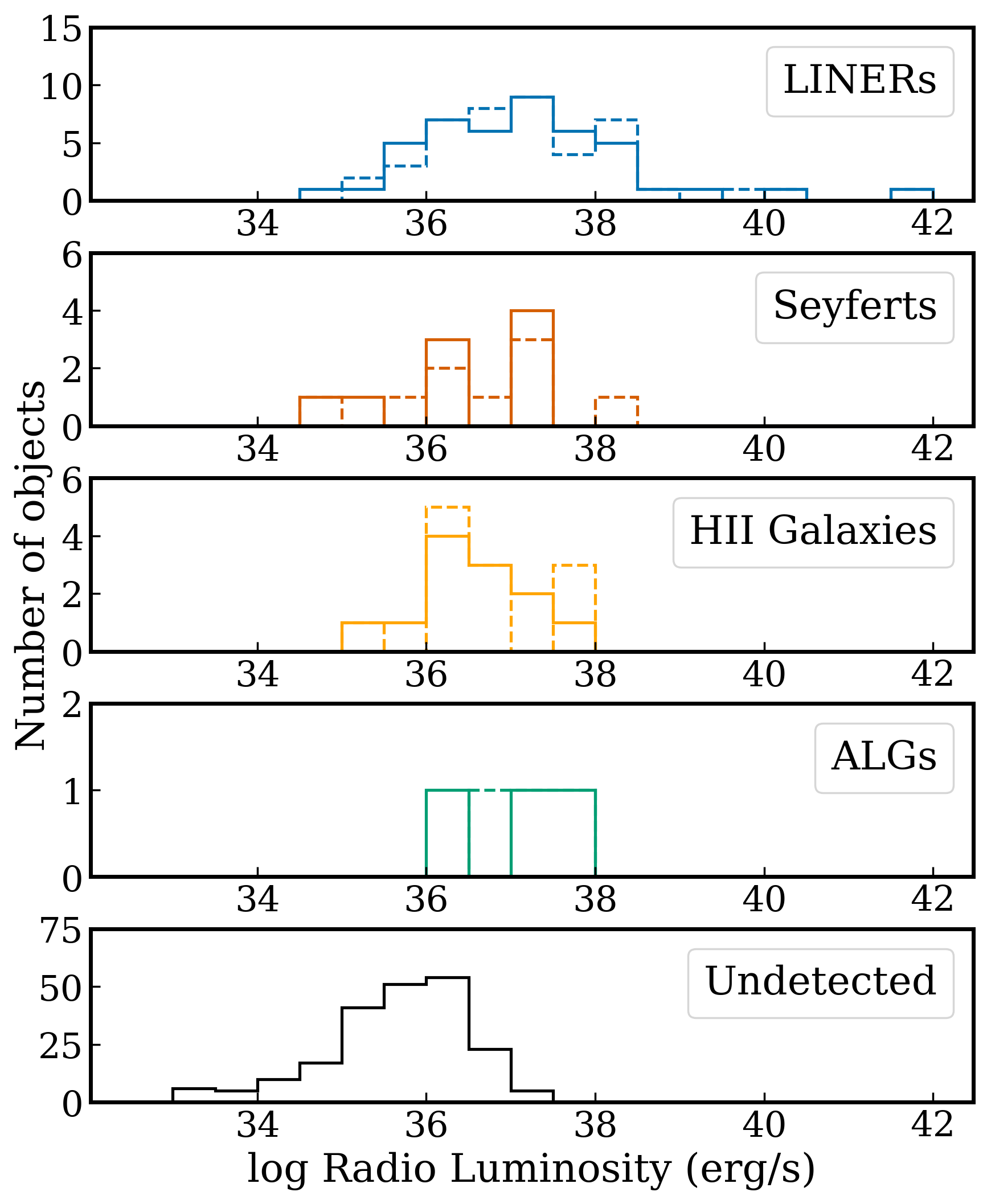}
   \caption{}
   \label{fig:histogramsbyAGNMORPHtypea} 
\end{subfigure}

\begin{subfigure}[b]{\columnwidth}
   \includegraphics[width=0.95\columnwidth]{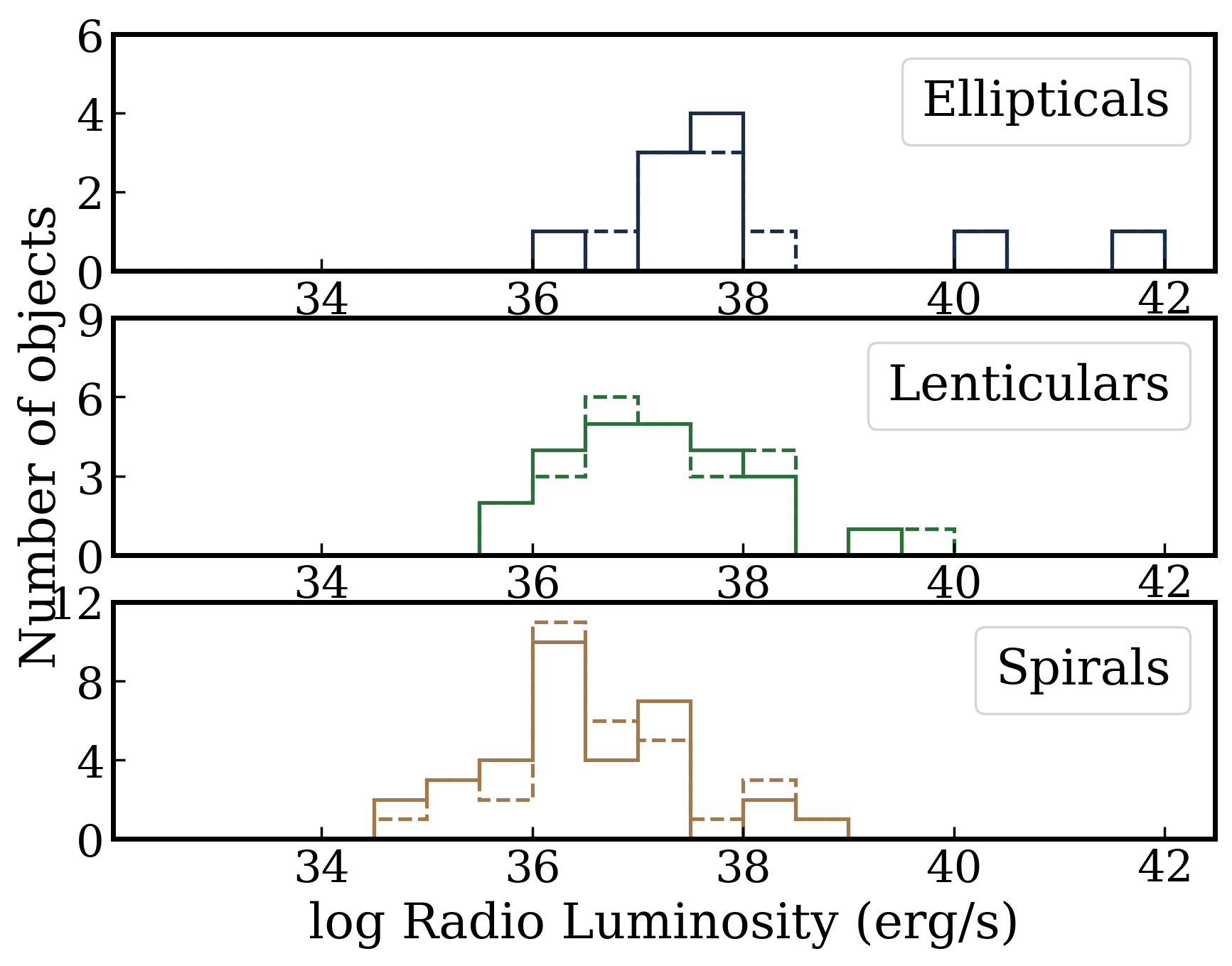}
   \caption{}
   \label{fig:histogramsbyAGNMORPHtypeb}
\end{subfigure}
\caption{The 5\,GHz radio luminosity distribution (erg s$^{-1}$) from the observations presented in this work per \textit{panel a: } optical nuclear class and \textit{panel b: } host morphological type. The radio core luminosity and the total radio luminosity distributions are outlined by the solid-line and the dashed-line histograms, respectively. The ``Undetected'' plot in panel a shows the 5$\sigma$ upper-limit radio luminosity distribution for the undetected sources. No irregular galaxies are detected and core-identified.}
    \label{fig:histogramsbyAGNMORPHtype}

\end{figure}

The detected, unidentified and undetected galaxies are summarised as a function of optical nuclear type and radio morphology for both the 1.5 and 5\,GHz samples in Table~\ref{tab:fraction2}. Our detection rates for all types are lower than in the 1.5\,GHz campaign. In particular, the detection fraction for `inactive' \ion{H}{II} galaxies and ALGs has fallen from 38/168 (1.5\,GHz) to 15/168 (5\,GHz). In contrast, the detection rate for the `active' Seyferts and LINERs has fallen less drastically from 68/112 to 53/112. 

Fig.~\ref{fig:histogramsbyAGNMORPHtype} shows the 5\,GHz luminosity distributions of the sample for both the peak radio core and total integrated radio luminosities, split by optical nuclear class (upper panels) and galaxy morphology type (lower panels). We compute uncensored and censored mean values\footnote{The uncensored mean is calculated using only detected sources, whereas the censored mean accounts for non-detections (upper limits) using survival analysis to provide a more representative statistical average of the entire population.} of the detected cores by AGN type. The `active' galaxies have the largest censored mean\footnote{The censored averages were computed using the \textsc{ASURV} package \citep{ASURV}.} values of $L_{\rm core} =$ 2.5$\times$10$^{35}$~erg~s$^{-1}$ for LINERs and $L_{\rm core} =$ 4.2$\times$10$^{35}$~erg~s$^{-1}$ for Seyferts. These values are similar to those found for the 1.5\,GHz sample \citep{BaldiLeMMINGs,BaldiLeMMINGs2,BaldiLeMMINGs3}. In contrast, the large number of non-detections in the `inactive' galaxies results in their censored mean values to be two orders of magnitude lower when compared to the 1.5\,GHz sample: $L_{\rm core} =$ 2.6$\times$10$^{33}$~erg~s$^{-1}$ for \ion{H}{II} galaxies and $L_{\rm core} =$ 3.0$\times$10$^{33}$~erg~s$^{-1}$ for ALGs. 
We also compute the mean values for the detected sources alone and found that detected Seyferts have the lowest mean luminosity $L_{\rm core} =$ 2.6$\times$10$^{36}$ erg s$^{-1}$, LINERs are clearly more luminous (mean $L_{\rm core} =$ 1.6$\times$10$^{37}$ erg s$^{-1}$), but their uncensored average is biased by the two sources $\geq$10$^{40}$ erg s$^{-1}$. The detected \ion{H}{II} galaxies have a similar mean to the Seyferts ($L_{\rm core} =$ 3.5$\times$10$^{36}$ erg s$^{-1}$), and the detected ALGs are brighter (mean: $L_{\rm core} =$ 1.0$\times$10$^{37}$ erg s$^{-1}$).

In terms of Hubble types, spirals make up the majority of detected sources but the smallest detection fraction due to the large number of spirals in the sample (33/189, 17 per cent). Only NGC~3690 is detected of the irregular galaxies, with a total LTG fraction of 16 per cent. In contrast, the ETGs have much higher detection fractions (34/76, 45 per cent): lenticulars are often detected (24/51, 47 per cent) as are ellipticals (10/25, 40 per cent). In contrast to the 1.5\,GHz sample where radio emission was detected in 33 per cent of spirals and 23 per cent of irregulars, the LTG detection fraction has fallen considerably, whereas the ETG detection fraction has only fallen from 51 per cent to 45 per cent. 
In general the ellipticals and lenticulars have the highest luminosities ranging from 10$^{36.4-41.9}$ erg s$^{-1}$ and 10$^{35.7-39.3}$ erg s$^{-1}$, respectively, with a censored mean luminosity of ETGs of 1.7$\times$10$^{35}$ erg s$^{-1}$. Spirals are detected at lower luminosities between 10$^{35-38.6}$ erg s$^{-1}$, and the LTG censored mean luminosity is 6.6$\times$10$^{33}$ erg s$^{-1}$.

\vspace{-1em}
\section{Discussion}
\label{sec:discussion}
The LeMMINGs sample is the deepest radio survey of the Palomar sample of nearby galaxies, down to $\sim$10$^{34}$ erg s$^{-1}$ ($\sim$10$^{17.5}$ W Hz$^{-1}$) at sub-arcsecond resolution, and the first to take an unbiased look at all galaxy morphological types for nuclear radio emission that may be associated with a LLAGN or SF. We detected radio nuclei within 1.5 arcsec of the \textit{Gaia} or best optical position from the literature in 106/280 galaxies in the 1.5\,GHz sample above a 3$\sigma$ detection limit of 0.24~mJy beam$^{-1}$, and here we identify radio emission 68/280 galaxies at 5\,GHz, above a 5$\sigma$ detection limit of 0.33~mJy beam$^{-1}$.  
We now discuss our findings, first comparing to previous radio surveys of the Palomar sample, then concentrating on the effect that using the \textit{Gaia} positions has on the detections, discussing the differences between the 1.5\,GHz and 5\,GHz samples, the detection -- or lack thereof -- of jets and evidence of SF in the sample and finally suggesting what is required for future surveys to ensure a better census of nuclear activity of the local Universe.

\vspace{-1em}
\subsection{Comparison with previous radio studies of the Palomar sample}
\label{sec:compstudies}

Multiple radio studies have been based on the Palomar sample at different resolutions (see Section~\ref{sec:Intro}). Many of these studies focused on the `active' galaxies such as LINERs or Seyferts, whereas the LeMMINGs sample is unbiased to the presence of an active SMBH. 
A complete analysis of all $\sim$200 `active' galaxies in the Palomar sample \citep{nagar05}, using data from the VLA and VLBA in the literature concluded that $\geq$50 per cent of all LINERs and Seyferts have a central accreting SMBH \citep[see also][]{nagar00,falcke00,filho00,HoUlvestad,nagar01,ulvestad01a,Nagar2002,anderson04,filho04,2018A&A...616A.152S}. \citet{nagar05} found that 38/39 (97 per cent) of sources with 15\,GHz VLA flux densities $>$2.7~mJy had sub-parsec-scale jets and high brightness temperature cores ($T_{\rm B} >$10$^7$K) when observed with the VLBA, indicating that they are genuine AGN. However, those with 15\,GHz VLA flux densities $<$2.7~mJy were not observed with VLBA and could also host genuine active SMBHs. 

In our 1.5\,GHz survey, compact radio cores were detected in 56/94 (60 per cent) of LINERs and 12/18 (67 per cent) of Seyferts. Here, our detection fractions are reduced to 44/94 (47 per cent) and 9/18 (50 per cent) for LINERs and Seyferts, respectively. LINERs show more extended radio structures than Seyferts, suggesting a higher capability of launching more collimated and more relativistic jets (e.g. \citealt{narayan08,nemmen14,Singh2015,BaldiLeMMINGs3}). These values for the `active' galaxies are consistent with the detection rates of high-resolution VLBA cores in previous Palomar-selected radio surveys \citep{nagar05}, but slightly less than the VLA studies \citep{HoReview}, suggesting that our intermediate-resolution observations between the VLA and VLBA have resolved out some of the sources. 
Nevertheless, the LINERs still typically have the brightest radio cores, and many of them have brightness temperatures $T_{\rm B} >$10$^6$K at VLBI resolution \citep{nagar05,Panessa2013}, or show evidence for parsec or kpc-scale jets. Therefore it is likely that all the detected LINERs host true LLAGN. As for the undetected LINERs, they may also hide LLAGN but deep VLBI surveys have not been able to detect radio emission: NGC~1058 and NGC~2685 \citep{Panessa2007A&A...467..519P} and NGC~404
\citep{Zsolt2014ApJ...791....2P}. In these cases the SMBHs may not be actively accreting, although deeper VLBI observations could still reveal radio emission for many of our undetected LINERs.

As for the detected Seyferts, they have lower average luminosity than LINERs by a factor $\sim$6 and are generally associated with compact radio cores in both the 1.5\,GHz and 5\,GHz samples. The brightness temperatures for these detected Seyferts range between $(0.6{-}0.8) \times 10^4$ K, so a higher resolution is needed to confirm a LLAGN. Fortunately, this has already been done with the VLBA/EVN, with detections reported for all our detected Seyferts \citep{falcke00,nagar05,Panessa2013}, except for NGC~5194 \citep{Bontempi2012MNRAS.426..588B,Cheng2025}, with high brightness temperature $T_{\rm B} >$10$^6$K. Furthermore, non-detected Seyferts in our sample have also been detected in deeper VLBI observations, e.g. NGC~4051 \citep{giroletti09} and NGC~4395 \citep{Yang2022MNRAS.514.6215Y}, though some remain notoriously elusive even at the 100~$\upmu$Jy level \citep[e.g., NGC~3185 and NGC~3941, see][]{Bontempi2012MNRAS.426..588B}. The evidence overwhelmingly suggests that Seyferts are unequivocally accreting SMBHs, able to launch sub-relativistic jets or outflows (e.g. \citealt{ulvestad05b,Middelberg2004}), and that they can be associated with non-thermal radio-emitting nuclei.

Of the LeMMINGs galaxies, 60 per cent are classified as `inactive': either \ion{H}{II} galaxies (140/280) or ALGs (28/280). Previous VLA observations of a statistically complete sample of 40 \ion{H}{II} galaxies detected no compact radio emission consistent with active SMBHs \citep{ulvestad02}. Consequently these sources were excluded from future samples \citep[e.g.,][]{nagar05}. In our 5\,GHz sample, we detected 12 \ion{H}{II} galaxies (9 per cent) and 3/28 ALGs (11 per cent) with an identified nuclear radio core. The detection fraction in the `inactive' galaxies has dropped by 23 per cent when compared to the 1.5\,GHz sample. 
The fall in detection rate at 5\,GHz of these `inactive' galaxies is driven by the \ion{H}{II} galaxies and suggests that some of the 1.5\,GHz \ion{H}{II} galaxy detections may have been contaminated with non-AGN processes. All fifteen of the sources detected below the 5$\sigma$ threshold (0.42~mJy beam$^{-1}$) in the 1.5\,GHz images are undetected at 5\,GHz, potentially showing that they are spurious detections \citep{BaldiLeMMINGs2}, although we cannot rule out steep spectra in these sources which would make them undetectable above this sensitivity limit (see also Section~\ref{sec:detectionLLAGNdiscussion}). For the twelve detected sources at 5\,GHz, nine have higher-resolution VLBA/VLBI data. NGC~2782, NGC~3245, NGC~3504 and NGC~3665 have high brightness temperature cores ($T_{\rm B} >$10$^6$K) and are probably LLAGN \citep{NGC37182007A&A...464..553K,Liuzzo2009A&A...505..509L,DellerMiddelberg,Cheng2025}. NGC~3690 is a tidal disruption event \citep{Mattila2005}. The diagnosis on the remaining four sources is uncertain, as their brightness temperatures are compatible with being caused by SN remnants \citep[NGC~2146 and NGC~4102,][]{2000A&A...358...95T,Cheng2025} or they are not detected with VLBI \citep[NGC~4041 and NGC~6946,][]{Cheng2025}. NGC~3504 may also host a weak LINER-type nucleus \citep{2000ApJ...530..688A} which adds additional uncertainty to the true number of LLAGN that may be present in \ion{H}{II} galaxies. Using our data and the literature, we can only be reasonably confident that 4/140 \ion{H}{II} galaxies hide a LLAGN, with a possible inclusion of the other two detected \ion{H}{II} galaxies. This would lead to an LLAGN detection fraction in \ion{H}{II} galaxies from 3$-$4 per cent. Deeper VLBI studies and ancillary multi-band information are still needed to confirm the presence of active SMBHs in these systems. 

In contrast to the \ion{H}{II} galaxies, 60 per cent of the ALGs detected at 1.5\,GHz remained detected at 5\,GHz, although this still corresponds to a small fraction of the number of ALGs in the overall sample (3/28, 11 per cent). The three detected ALGs in our 5\,GHz sample (NGC~507, NGC~2300 and NGC~3348) were also recently detected using VLBI \citep{Cheng2025} with high brightness temperature cores ($T_{\rm B} >$10$^{6.5}$K) and thus these are all plausible AGN candidates, though further work is required to rule out other non-thermal emission mechanisms. In the previous LeMMINGs work, we found that ALGs in ellipticals have characteristics indistinguishable from LINERs \citep{BaldiLeMMINGs3,X-rayLeMMINGs} and their lack of optical detections may make them similar to those found in local radio-loud AGN \citep{baldi10b,heckman14}. However, their peak intensities in our 5\,GHz sample are $\sim$1~mJy beam$^{-1}$ with luminosities between $(0.2{-}3.2)\times$10$^{37}$ erg s$^{-1}$. Deeper observations of ALGs are required to fully understand the nature of nuclear radio emission in these objects and rule out stellar processes \citep{PaudelYoon,capetti22,stasinska25}. 

The high radio detection of nuclei within the LeMMINGs survey highlights a significant prevalence of accretion-powered systems across various host galaxy types. Considering the 1.5\,GHz and 5\,GHz datasets together, we identify a total of 79 radio-detected nuclei (with signatures of SMBH activity or jets) within the sample of 280 galaxies. This census is primarily composed of 58 LINERs -- consisting of 56 previously identified at 1.5\,GHz plus an additional 2 detected at 5\,GHz -- alongside 12 Seyfert galaxies, 7 jetted \ion{H}{II} galaxies, and 2 jetted ALGs. This results indicates that a approximately one third of the galaxies (79/280, ETGs in particular) in the local Universe hosts an active SMBH, capable of producing significant radio emission with flux densities exceeding $\sim$0.3 mJy. This higher detection fraction of radio LLAGN in our LeMMINGs campaign than previous radio surveys ($\sim$10 and $\sim$20 per cent with FIRST and LOFAR surveys with mJy sensitivities for z$\sim$ 0.1-0.3, e.g. \citealt{Best05,baldi10,sabater19}) emphasizes that high-resolution, sensitive radio imaging is essential in detecting pc-scale emission at the faintest levels.

\vspace{-1em}
\subsection{Positional core offsets among optical, 1.5 and 5 GHz data}
\label{sec:Gaiaoffsetsdiscussion}

\begin{figure}
\centering
   \begin{subfigure}[b]{\columnwidth}
   \includegraphics[width=0.90\columnwidth]{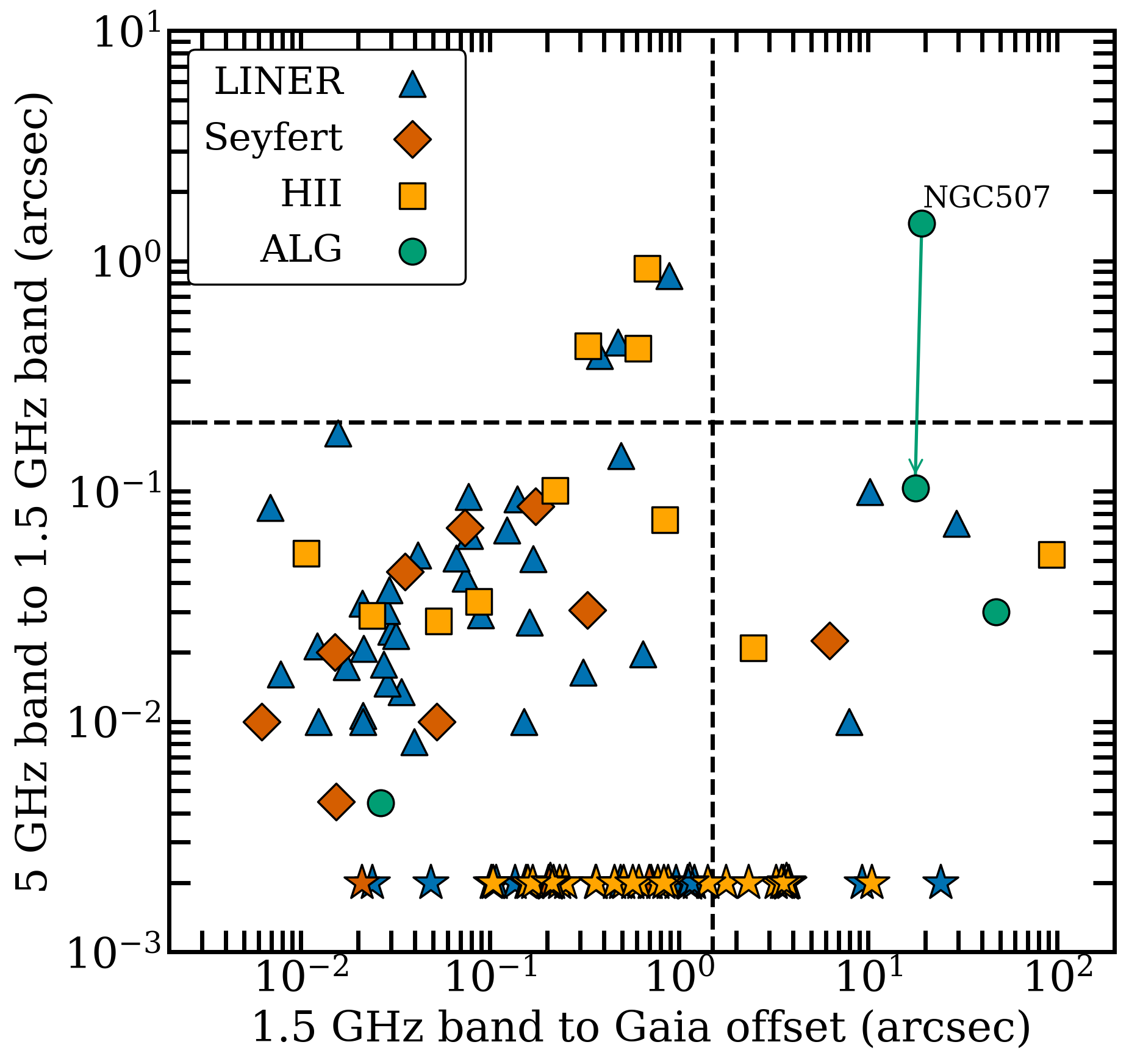}
   \caption{}
   \label{fig:Gaiaposoffseta} 
\end{subfigure}

\begin{subfigure}[b]{\columnwidth}
   \includegraphics[width=0.90\columnwidth]{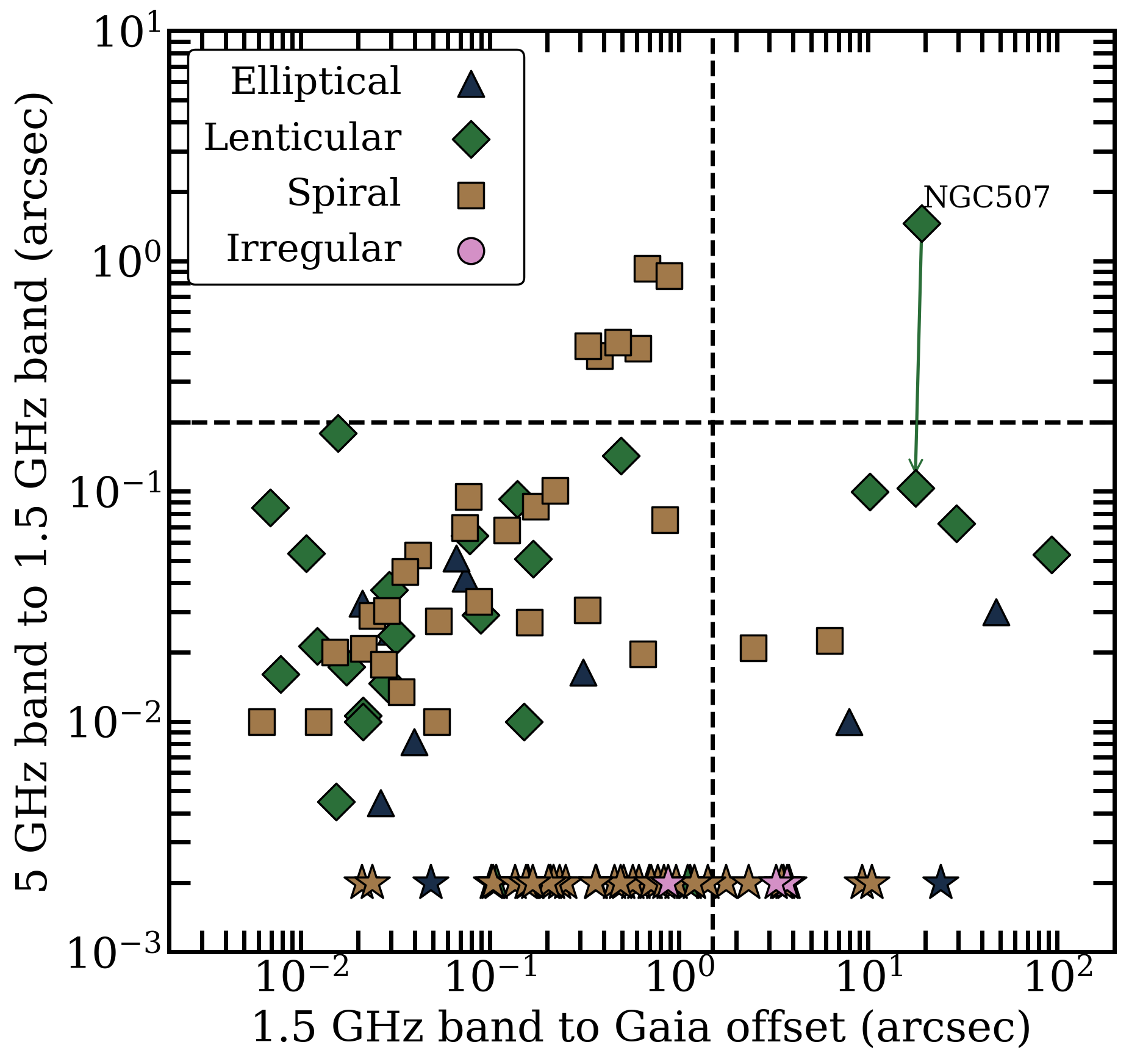}
   \caption{}
   \label{fig:Gaiaposoffsetb}
\end{subfigure}
\caption{Positional offsets in arcseconds between the detected 1.5\,GHz LeMMINGs sources and the \textit{Gaia} optical position on the $x$-axis, and between the detected 5\,GHz LeMMINGs sources and the 1.5\,GHz LeMMINGs positions of those sources on the $y$-axis. The detected sources are split by colour and shape shown in the legend of each plot. The sources that are undetected in the 5\,GHz data -- but detected in the 1.5\,GHz data -- are shown at the bottom of each plot as stars, at a $y$-axis value of 0.002 arcsec. The \textit{top} plot shows the sources by optical nuclear classes, whereas the \textit{bottom} plot separates them by galaxy morphological types. 
    The vertical dashed line refers to an offset distance of 1.5 arcsec, i.e. the search region used in the 1.5\,GHz data for nuclear radio emission. 
    The horizontal dashed line is plotted at an offset of 0.2 arcseconds, i.e. the average \textit{e}-MERLIN synthesized beam size of the 1.5\,GHz data. The position of NGC~507 is marked in both plots, with the arrow indicating the change in position of this plot once taking into account the new 1.5\,GHz data reduction position of this source, i.e. it moves below the horizontal line (see Section~\ref{sec:Gaiaoffsetsdiscussion}).}
    \label{fig:Gaiaposoffset}

\end{figure}

In Fig.~\ref{fig:Gaiaposoffset} we show the offset between the 1.5\,GHz positions \citep{BaldiLeMMINGs,BaldiLeMMINGs2} and the \textit{Gaia} position (on the $x$-axis) and the 5\,GHz position (on the $y$-axis), split by the optical nuclear type (\textit{top} plot) and the galaxy morphological type (\textit{bottom} plot). We also mark the 1.5 arcsecond search region from the 1.5\,GHz data for radio emission as a vertical dashed line and the \textit{e}-MERLIN synthesized beam size at 1.5\,GHz (0.2 arcseconds) as a horizontal dashed line. Detected sources to the right of the vertical line (NGC~507, NGC~2146, NGC~2300, NGC~2655, NGC~3665, NGC~4278, NGC~5194 and NGC~5866) are those which the \textit{Gaia} position disagrees with the detected radio position by more than 1.5 arcseconds due to the Simbad position being used. There are also ten sources (NGC~147, NGC~1560, UGC~4028, NGC~2832, NGC~7080, NGC~3448, NGC~3898, NGC~4100, NGC~4217, NGC~4244) with larger \textit{Gaia} offset that are not detected at 5\,GHz. These offsets between \textit{Gaia} and the radio data could be caused by misidentification of the optical galaxy core due to a foreground star. For example, in the case of NGC~2655 (see further discussion of the radio morphology of this source in Section~\ref{sec:largescalejets}), the \textit{Gaia} position appears to align with a star that is not related to NGC~2655. For all sources where there was a large discrepancy between the Simbad and \textit{Gaia} positions, we made radio images at both positions and found no additional radio sources, indicating that we have likely not missed any radio-emitting nuclei in the sample.

The \textit{top} plot in Fig.~\ref{fig:Gaiaposoffset} does not show a relation between the nuclear types and the number of detected sources with a \textit{Gaia} positional offset of $\geq$1.5 arcseconds. However, of the eighteen sources in this region, nine (50 per cent) are \ion{H}{II} galaxies and only two of these are detected in the 5\,GHz data (NGC~2146 and NGC~3665). Furthermore, many (7/9) of these \ion{H}{II} galaxies are spirals, all  highly-inclined or edge-on, which may explain why the \textit{Gaia} position and the optical positions used in the 1.5\,GHz dataset are discrepant. In the \textit{bottom} plot of Fig.~\ref{fig:Gaiaposoffset}, half (4/8, NGC~507, NGC~2655, NGC~3665 and NGC~5866) of the detected sources on the right-hand-side of the plot are lenticular galaxies. Several of these host known LLAGN, i.e., NGC~2655, NGC~3665, may have had galaxy mergers in the past \citep[e.g.,][]{NGC2655merger2016AAS...22724012R}, which may point to a possible offset between the active SMBH and the galaxy centre. Yet, the low number of data points does not allow statistical confidence in this assertion.

\vspace{-1em}
\subsubsection{Radio detections between the 5 and 1.5\,GHz data}

Around half of the 1.5\,GHz detections (125/280) are not detected in the 5\,GHz data (68/280). Where sources were detected at 1.5\,GHz and not at 5\,GHz (57 sources), we also searched the region around the 1.5\,GHz position for significant radio emission. As expected, we did not detect any radio emission in these sources as the majority of the LeMMINGs sources (244/280, 80 per cent) the \textit{Gaia} optical positions agree with the optical position used in the 1.5\,GHz campaign: this indicates that the radio detections are independent of our choice of optical position. 

Most of the sources have matched 1.5 and 5-GHz radio positions, with typical differences consistent with the beam of the 1.5 GHz images, 0.2 arcsec (symbols falling below the horizontal line in Fig~\ref{fig:Gaiaposoffset}). The median positional offset between the two samples is $\leq$30 mas, when the discrepant sources are removed. This close consistency is expected, as flat-spectrum radio cores of LLAGN could show small frequency-based offset due to jet/outflow/SF resolved structures shifting the core position \citep[e.g.,][]{Porcas}. The self-calibration procedure could potentially also shift the positions but only by a fraction of the synthesized beam. As an example, NGC~1167 shows a core with an extension to the south-east in the 1.5\,GHz image, but the exact position in the 5\,GHz image is $\sim$0.18-arcsec along this jet direction away from the 1.5\,GHz core position.

\begin{figure}
    \centering
    \includegraphics[width=0.95\columnwidth]{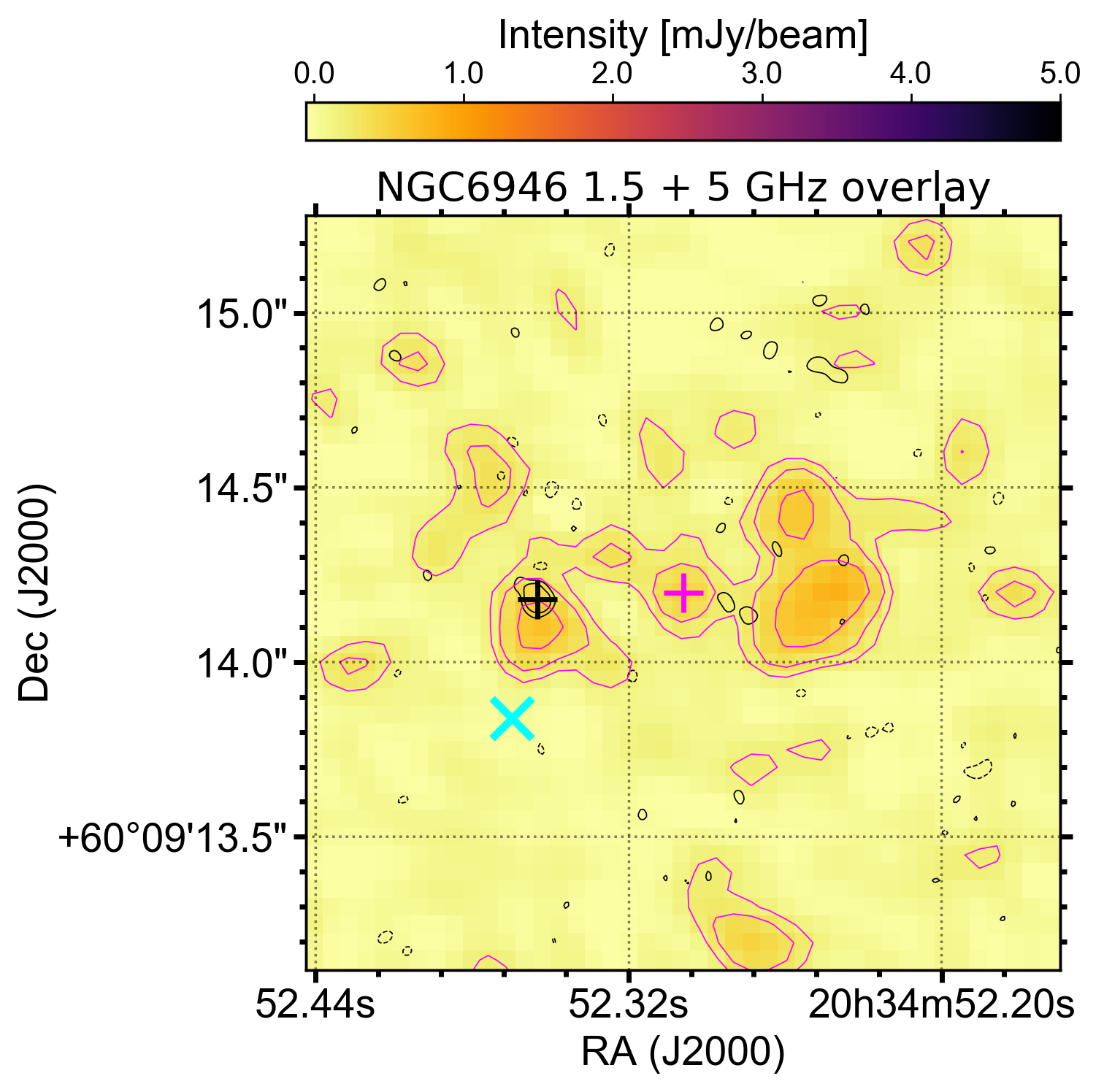}
    \caption{Image of NGC~6946 using 1.5\,GHz and 5\,GHz \textit{e}-MERLIN data. The false colour background image is the 1.5\,GHz image presented in \citet{BaldiLeMMINGs} with magenta contours representing the $-$3, 3, 5, 9, 16 $\sigma$ with 1$\sigma$ equal to 57~$\upmu$Jy beam$^{-1}$. The black contours represent the 5\,GHz detection of NGC~6946 discussed in this work, with the contours used: $-$3, 3, 5, 9 $\times$ image r.m.s. sensitivity, which is 69~$\upmu$Jy beam$^{-1}$. Negative contours are dashed. The cyan `$\times$' denotes the \textit{Gaia} position of NGC~6946, the magenta `+' shows the 1.5\,GHz source position and the black `+' shows the 5\,GHz source position from this work. The complex nature of the nuclear region renders the AGN position ambiguous.}
    \label{fig:NGC6946}
\end{figure}

Seven sources with offsets larger than the 1.5-GHz \textit{e}-MERLIN beam size, e.g., those that fall above the horizontal line in Fig.~\ref{fig:Gaiaposoffset} (NGC~507, NGC~2841, NGC~4041, NGC~4102, NGC~4736, NGC~6946 and NGC~7217), have optical positions that do not fall within the 5\,GHz beam area either. Overlays of the 1.5\,GHz and 5\,GHz data for these seven sources are presented in the online supplementary material. 
For these sources except NGC\,507, the radio emission in the 1.5\,GHz data shows extended and complex morphology (see NGC~6946 as an example in Fig.~\ref{fig:NGC6946}). All are hosted in a spiral galaxy. The difficulty of identifying the core position in such entangled radio structures, between a putative jet and SF components corroborates the need for multi-band observations of galactic nuclei to pinpoint the genuine position of the central SMBHs.

\begin{figure}
    \centering
    \includegraphics[width=0.95\columnwidth]{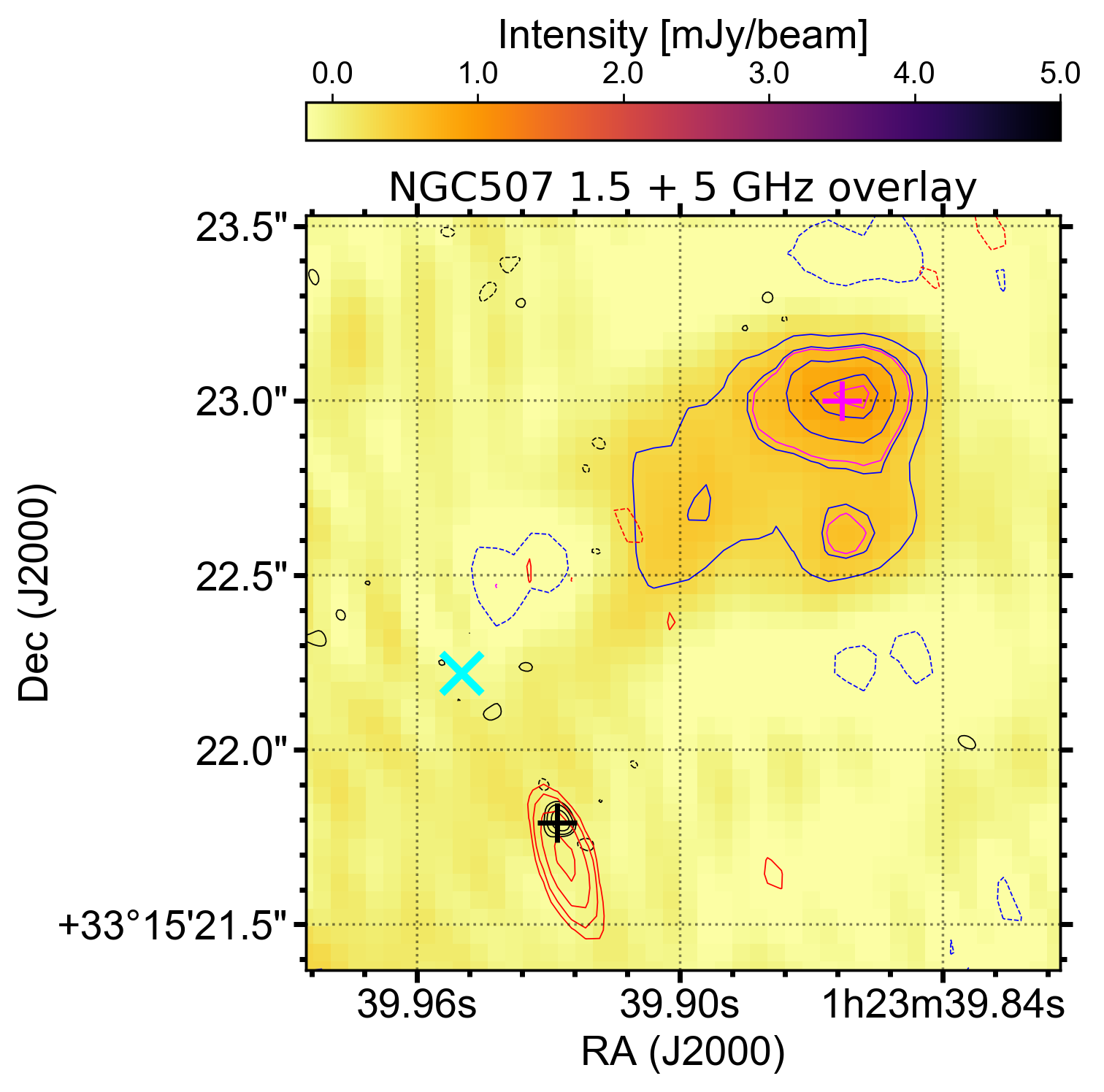}
    \caption{Image of NGC~507 using 1.5\,GHz and 5\,GHz \textit{e}-MERLIN data. The false colour background image is the 1.5\,GHz image presented in \citet{BaldiLeMMINGs} with magenta contours representing the $-$3, 3, 5 $\sigma$ with 1$\sigma$ equal to 57~$\upmu$Jy beam$^{-1}$. The blue contours represent the same contour scheme used in \citet{BaldiLeMMINGs}: 0.34$\times (-$1, 1, 1.5, 2, 2.5)\,mJy~beam$^{-1}$. The black contours represent the 5\,GHz detection of NGC~507 discussed in this work, with the contours used: $-$3, 3, 5, 9 $\times$ image r.m.s. sensitivity, which is 82 $\upmu$Jy beam$^{-1}$. The red contours ($-$3, 3, 5, 9, 16 $\times$ image r.m.s. sensitivity, which is 93 $\upmu$Jy beam$^{-1}$) show the re-processed LeMMINGs 1.5\,GHz data, using the same technique outlined in this work. Negative contours are dashed. The difference in position between the previous analysis and new analysis leads to the vertical drop in Fig.~\ref{fig:Gaiaposoffset}. The cyan `$\times$' denotes the Simbad position of NGC~507. The magenta `+' shows the 1.5\,GHz source position and the black `+' shows the 5\,GHz source position from this work.}
    \label{fig:NGC507}
\end{figure}

For NGC~507, a re-analysis of the 1.5\,GHz data\footnote{This work is part of a re-analysis of the 1.5\,GHz LeMMINGs campaign which is currently underway using \textsc{CASA} and the same self-calibration method as described in Section~\ref{sec:selfcal}. } shows that the previously reported position is incorrect. Following this re-analysis, the 1.5\,GHz and 5\,GHz positions agree with one another (see Fig.~\ref{fig:NGC507}. The new radio position also agrees with the Simbad position for this source to within $\sim$0.5 arcsec. This change is reflected in Fig.~\ref{fig:Gaiaposoffset} with an arrow, showing the new position on this plot, reconciling the two radio positions of this source in our datasets.

\vspace{-1em}
\subsubsection{Sources detected at 5\,GHz but not at 1.5\,GHz}

In two cases we detect a source at 5\,GHz that was undetected at 1.5\,GHz: NGC~3642 and NGC~4138. We note that the \textit{Gaia} position and literature position disagree by $\leq$0.3 arcseconds for both these sources, indicating that the lack of 1.5\,GHz detection is not driven by the previous choice of optical position.

NGC~3642 is a LINER spiral galaxy, which has a peak intensity of 0.55$\pm$0.06~mJy beam$^{-1}$ in our 5\,GHz data, but a 3$\sigma$ upper limit of 0.261~mJy beam$^{-1}$ in the 1.5\,GHz data. The only previous VLA survey to detect this source was conducted at 15\,GHz and matches closely our 1.5\,GHz resolution (0.15--0.2 arcsec, \citealt{2018A&A...616A.152S}). The source peak intensity at 15\,GHz was 0.78$\pm$0.01~mJy beam$^{-1}$ and the positions of the 5\,GHz and 15\,GHz data agree. If the source has not varied between observations, then the radio spectral index is slightly inverted: $\alpha =$ 0.32, which would suggest a detection of $\sim$0.35~mJy may have been possible at 1.5\,GHz. The inverted radio spectral index could be due to free-free/synchrotron absorption in the nuclear region, though X-ray spectral fits suggest absorption in addition to the Galactic absorbing column is not required \citep{X-rayLeMMINGs}. The origin of such an inverted spectrum could be a sign of a new born radio jet, like in gigahertz peaked spectrum sources \citep[e.g.,][]{Yu}. Since the observations are not simultaneous, caution must be taken to first rule out a transient or variable nature of its nuclear emission (e.g. \citealt{padovani16,odea21,chen24}).

NGC~4138 is a LINER lenticular galaxy, which was detected in our 5\,GHz data at 0.53$\pm$0.03~mJy beam$^{-1}$, but was undetected at 1.5\,GHz at a 3$\sigma$ threshold of 0.375~mJy beam$^{-1}$. It was detected at a peak intensity of 1.5~mJy beam$^{-1}$ by \citet{nagar05} at 15\,GHz. Interestingly, this source is also detected in VLBI observations at 5\,GHz at 0.74~mJy and 1.5\,GHz at 1.0~mJy \citep{Bontempi2012MNRAS.426..588B}, indicating a radio spectrum $\alpha \sim -$0.3. A second fainter component was also detected in the VLBI images to the west of the core. Possible nuclear variability \citep[in X-ray, ][see also Section~\ref{sec:detectionLLAGNdiscussion}]{hernandez17} and the resolved-out structures at different resolutions could accommodate the non-detection of this source in our shallow 1.5\,GHz image.

\subsubsection{Radio Spectral Indices between the 1.5\,GHz and 5\,GHz data}
\label{sec:specindex}

\begin{figure}
    \centering
    \includegraphics[width=0.95\columnwidth]{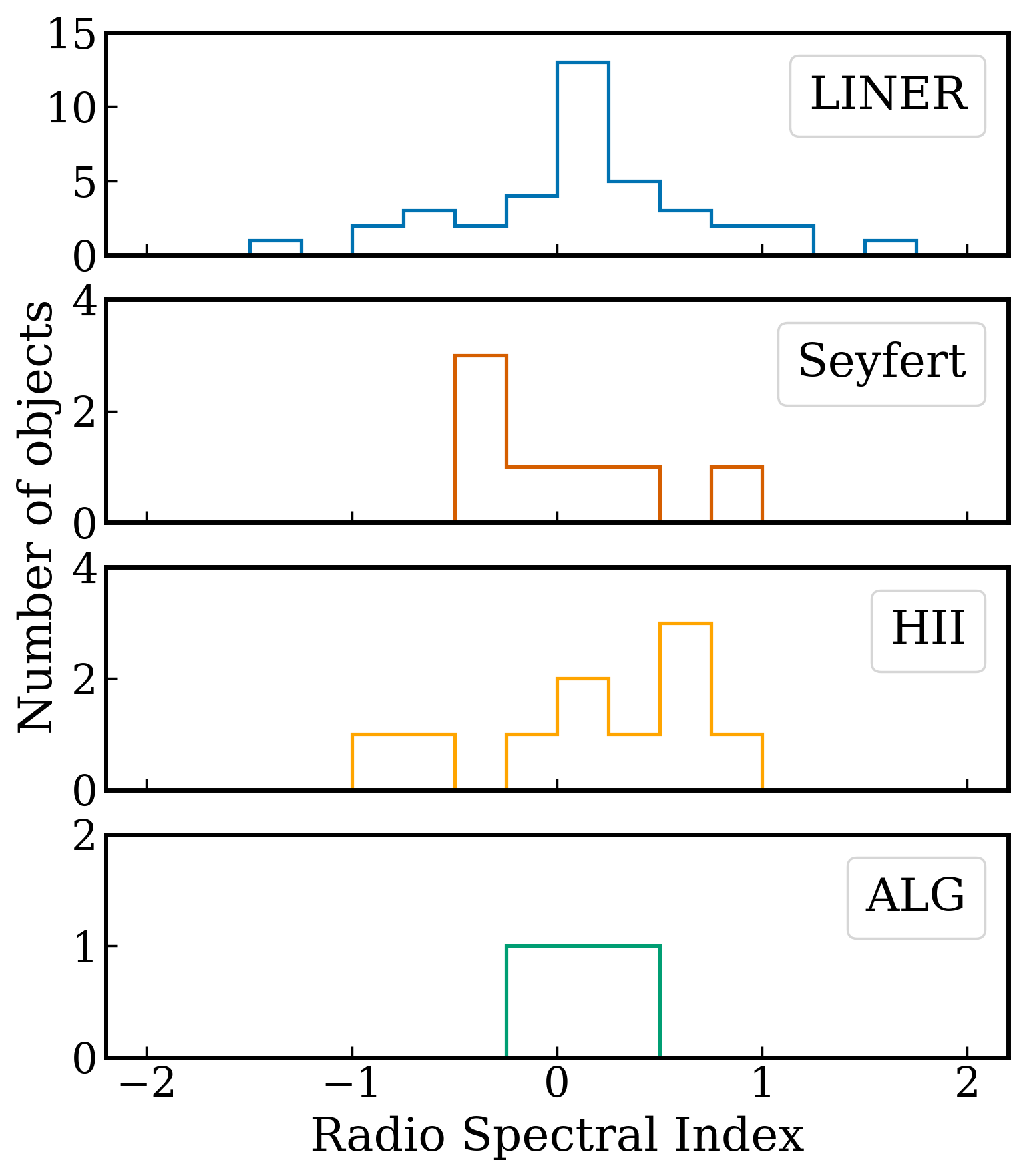}
    \caption{Radio spectral index $\alpha$ by AGN type from detections in both the 1.5\,GHz full-resolution data and 5\,GHz tapered resolution data (see Sect.~\ref{sec:specindex} for details). The bin width is 0.25 for all subplots. A better analysis of $\alpha$ will be subject of a future work.}
    \label{fig:specindex}
\end{figure}

The radio spectral index can be a useful tool in discriminating between radio emission mechanisms in radio-quiet/loud AGN \citep{Panessa2019,zajacek19}. Fig.~\ref{fig:specindex} displays the non-simultaneous radio spectral index distribution per AGN type for all sources except NGC~3690 detected in the 5\,GHz tapered-resolution \emerlin{} data, approximately matched in beam size with the 1.5\,GHz data presented in \citet[][]{BaldiLeMMINGs,BaldiLeMMINGs2} ($\sim$0.2 arcsecond). NGC~3690 is removed due to the known variable nature of the source. We use the total flux measurements of the core component in both images. If either observation resulted in an upper limit, we discount the source to ensure reliable spectral index measurements. We find a range of radio spectral index values from +1.5 (NGC~3079) to $-$1.3 (NGC~1167). The average radio spectra per class are: $\alpha$ = 0.13 (LINERs), $-$0.01 (Seyferts), 0.18 (\ion{H}{II} galaxies) and 0.17 (ALGs). Given that many of the 5\,GHz objects are compact, and show approximately flat radio spectra between these two bands, we suggest that most sources detected in both samples must be compact radio cores associated with an LLAGN.  Previous observations of the Palomar sample have shown an analogous broad range of spectral indices ranging between $-$1 $\lesssim \alpha \lesssim$ +1 \citep{nagar01,ulvestad01a,Nagar2002,nagar05}.

It is plausible that different mechanisms are responsible for the larger range in these spectral index values, with positive values potentially due to an advection dominated accretion flow (ADAF) \citep{falcke01}, young radio activity from the nucleus and low frequency absorption \citep{odea21}. The negative values could arise from contaminating jet emission. A good example of the latter issue is NGC\,4151 (see Section~\ref{sec:largescalejets}) where the flux density of the AGN core at full-resolution at 5\,GHz is one tenth of the nearby first jetted component, but they are unresolvable in the tapered resolution 5\,GHz image. In addition, intrinsic source variability could bias the radio spectrum but we cannot account for that with these data. 

A more detailed analysis is required to match the \textit{uv}-coverage of the images between the 1.5\,GHz and 5\,GHz images and ensure the synthesized beams are identical. This issue may account why NGC~1167, which has large-scale radio jet emission, has such a steep radio spectrum as the tapered 5\,GHz image has a synthesized beam of 0$\farcs$11$\times$0$\farcs$07, whereas the 1.5\,GHz image has a synthesized beam of 0$\farcs$2$\times$0$\farcs$2. Such an analysis, with each galaxy being matched exactly in \textit{uv}-range and synthesized beam, will be performed in a forthcoming work to ensure a more precise comparison between the datasets and provide a more reliable spectral index measurement.

\vspace{-1em}
\subsection{The presence of jets in the 5 GHz LeMMINGs sample}
\label{sec:jetteddiscussion}

One of the key benefits in observing at 5\,GHz is the higher resolution (50\,mas) achieved when compared to the 1.5\,GHz (150\,mas) observations, which probes four times larger linear scales. The improved resolution and increased \textit{uv}-filtering at higher frequencies resolve-out the extended radio emission, providing a more accurate description of the nuclear activity, without contamination from the jets, or from non-AGN related processes like SF. 

Of the 106 detected sources in the 1.5\,GHz sample, we classified 64 as core/core-jet, with the rest (42) having either a one-sided jet, triple, double-lobed or complete jet structure, i.e. 60 per cent of the sources had an unambiguous unresolved core. In the 5\,GHz sample, we identify nuclear radio emission in 68 targets and 15 of those show jetted structures (15/68, 22 per cent). The lower fraction of observed jet structures in the 5\,GHz sample confirms that the higher frequency data have been successful in isolating nuclear emission, albeit with far fewer detections: 38 per cent at 1.5\,GHz compared to 22 per cent at 5\,GHz. For the 15 sources with resolved structures, most show only an extension to one side (10/15, 67 per cent) enabling an unambiguous core detection and parameter extraction. The difficulty in extracting a core flux density for the triple and complex structures is alleviated by the improved resolution. Thus, the sample is relatively `clean' in terms of providing bona-fide core detections, which is especially important when comparing radio emission to other tracers of activity, such as optical or X-ray emission. 

The 5\,GHz data can be used to clearly identify the radio `cores' of sources even those with detected jet emission. An excellent example is NGC\,2639, which has shown episodic AGN activity with a re-oriented jet \citep{Rao2023MNRAS.524.1615R}. Our 5\,GHz images probe the $\sim$100\,pc intermediate scales between the 360\,pc scale east-west jet observed in VLA 5\,GHz data (beam size $\approx$0.4 arcsec) and the 3 parsec-scale jet at PA = 130$^{\circ}$ observed by the VLBA with a resolution of $\approx$7\,mas \citep[see Fig.~2 of ][]{Rao2023MNRAS.524.1615R}. Our 5-GHz \textit{e}-MERLIN observation detected three significant ($>$5$\sigma$) components aligned east-west with the brightest component aligning to within 40\,mas of the brightest component in the VLBA data. We also resolved the protrusion to the south-east of the brightest component, which is clearly observed in the VLBA images.

Some compact sources labelled with the compact `A' morphology or the identified cores in extended structures may hide pc-scale jets. In fact, we know this is the case in several of the brighter sources above 100~mJy beam$^{-1}$ which  all have jets at VLBI resolution:   NGC~1167 \citep{suma19}, NGC\,315 \citep{ParkNGC315}, NGC~1275 \citep{Giovannini2018NatAs, Paraschos2022}, NGC~3079 \citep{Irwin1998}, NGC~3998 \citep{Yan2025} and NGC~4278 \citep{2004evn..conf...81G}.  There are also examples of VLBI detections of resolved jet structures in some of the fainter objects, such as NGC~2146 and NGC~3884 which show additional components when observed with the EVN \citep{Cheng2025}.

\vspace{-1em}
\subsubsection{Large-scale jets}
\label{sec:largescalejets}

We do not detect the extended jets at 5 GHz in the brightest radio AGN which are known to display large-scale jetted structures, see e.g., NGC~315 \citep{ParkNGC315}, NGC~1275 \citep{Gendron2021}, NGC~3079 \citep{Middelberg2007,Sebastian}, NGC~3998, and NGC~4278 \citep{WrobelHeeschen}. This may be due in part to the \textit{uv}-filtering of the 5\,GHz data as the largest angular scale probed by \textit{e}-MERLIN is $\approx$1 arcsecond\footnote{The largest angular scale is calculated by using the shortest 12~km baseline of \textit{e}-MERLIN from Mk2 to Pickmere.}. The `snap-shot' imaging mode used in this study provides a spread-out but sparsely-sampled \textit{uv}-coverage, which circularises the synthesized beam, but is not sufficient for detecting faint structures, or providing high dynamic range images. 

\begin{figure}
\centering
   \begin{subfigure}[b]{\columnwidth}
   \includegraphics[width=0.95\columnwidth]{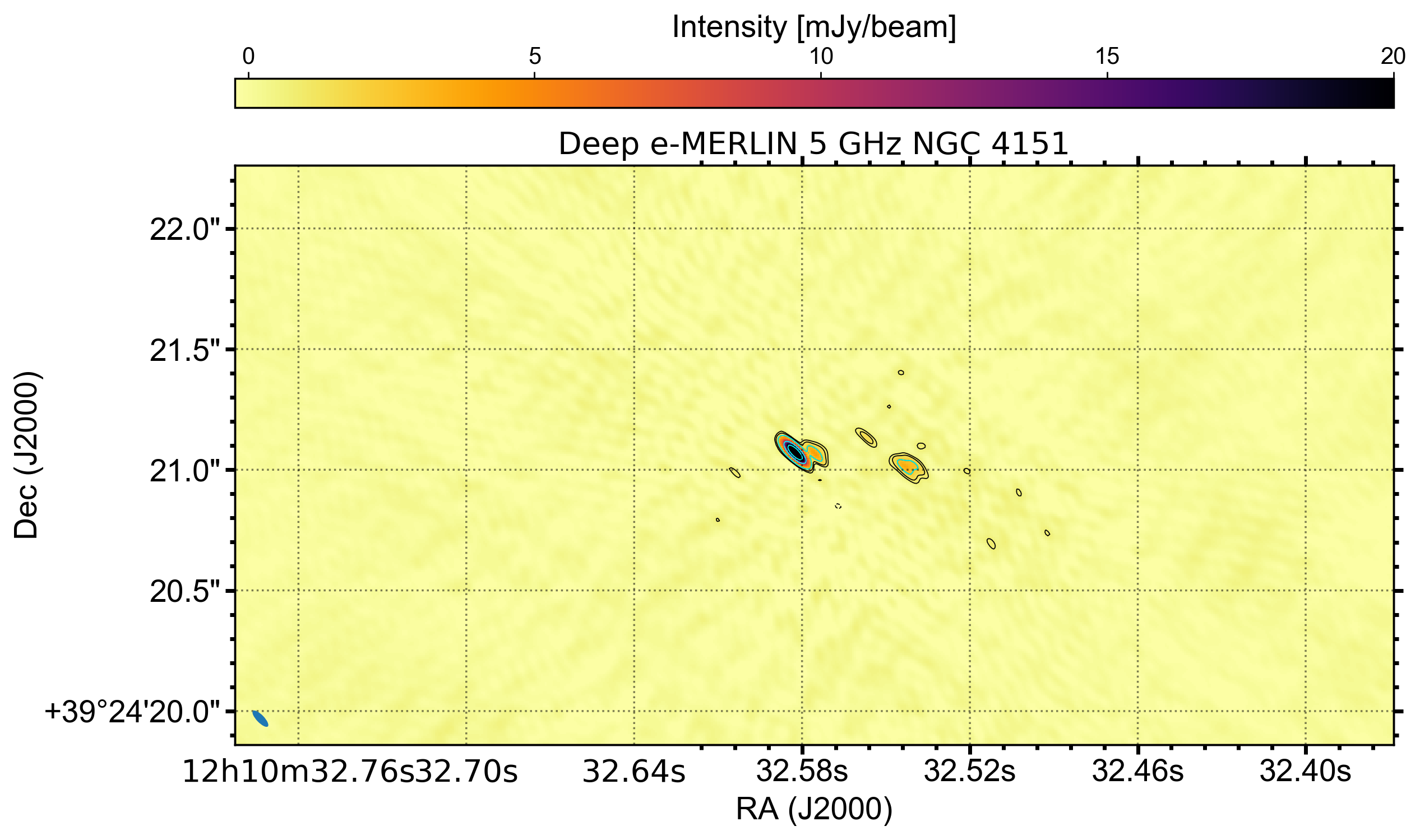} 
   \caption{}
   \label{fig:NGC4151compa} 
\end{subfigure}

\begin{subfigure}[b]{\columnwidth}
   \includegraphics[width=0.95\columnwidth]{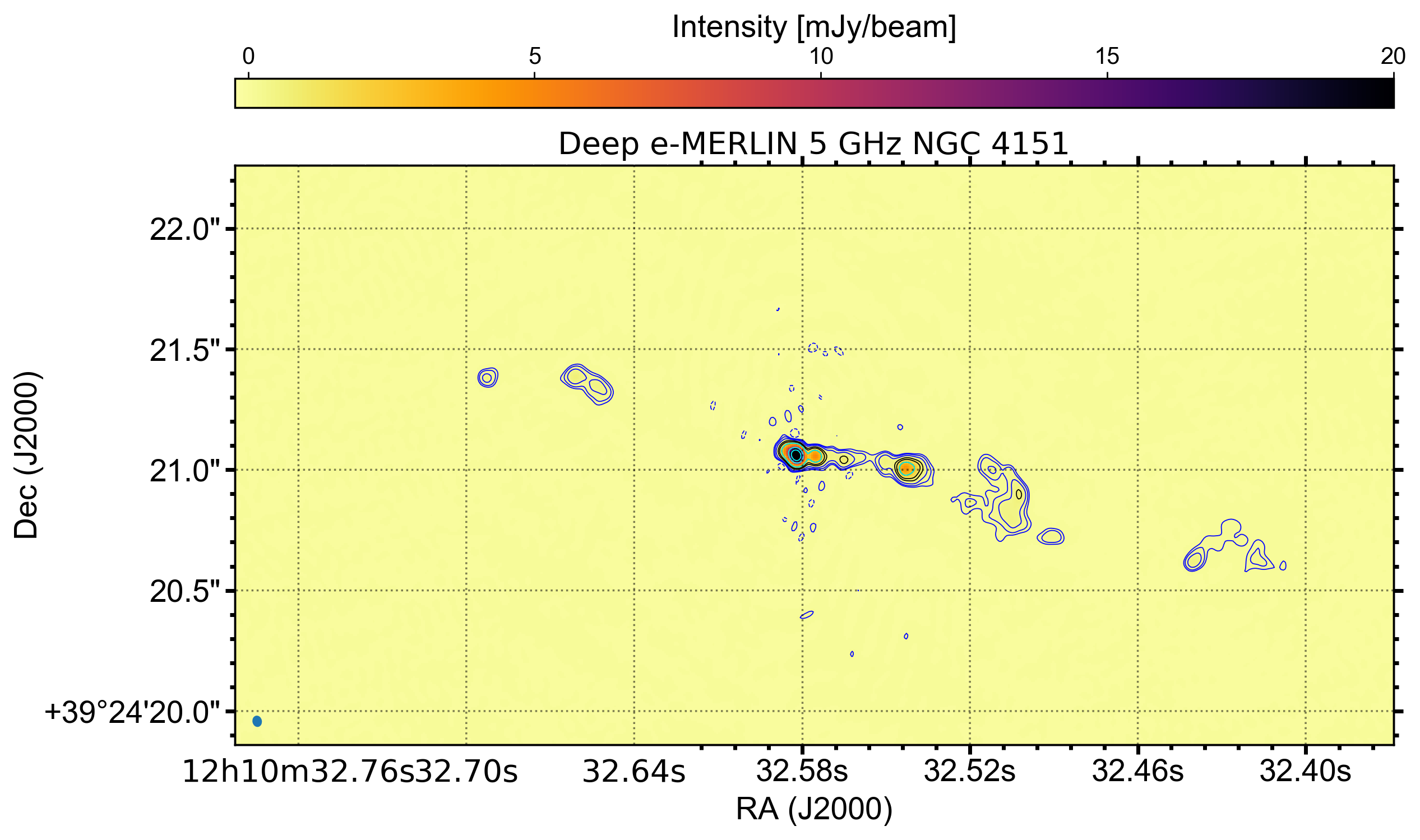}
   \caption{}
   \label{fig:NGC4151compb}
\end{subfigure}
\caption{Images of NGC\,4151 in \textit{panel a:} LeMMINGs \textit{e}-MERLIN 5\,GHz data presented here (r.m.s. sensitivity 235~$\upmu$Jy beam$^{-1}$), \textit{panel b:} deep 5\,GHz \textit{e}-MERLIN data obtained in 2018 \citep[r.m.s. sensitivity 30~$\upmu$Jy beam$^{-1}$,][]{Williams4151}. In both images, the black and cyan contours are plotted identically in the LeMMINGs and deep data: black contours plotted at 235 $\times$ ($-$3, 3, 5)~$\upmu$Jy beam$^{-1}$ and cyan contours at 235 $\times$ (10, 50, 100)~$\upmu$Jy beam$^{-1}$. In the deep image, additional blue contours are plotted 30$\times$ ($-$3, 3, 5, 10)~$\upmu$Jy beam$^{-1}$. Negative contours are dashed. The synthesized beam of each image is shown in the bottom left-hand corner of the image. }
    \label{fig:NGC4151comp}

\end{figure}

To illustrate this point further, we use the example of the archetypical Seyfert galaxy NGC\,4151 \citep{Kaiser2000}. It has a bi-polar jet structure at 5\,GHz with `deep' \textit{e}-MERLIN observations obtained a year prior to the LeMMINGs observations presented in this work \citep{Williams4151_2}. We plot both the LeMMINGs and `deep' data in Fig.~\ref{fig:NGC4151comp}. The `deep' data show several additional components extended over 3.5 arcsec in length. Our 5-GHz data only display three components extended over 0.75-arcsec near the optical centre of the galaxy. The snapshot strategy adopted for the LeMMINGs campaign led to partial $uv$ coverage and reduced sensitivity to the large-scale extent of the sources, particularly at 5 GHz.

However, two sources (NGC~3516 and NGC~4036) show larger-scale jets further away than the central cut out regions. NGC~3516 is a Seyfert galaxy with a known jet \citep{Miyaji1992ApJ...385..137M,ghosh2025} extending to the north and gradually bending eastward, with a full extent of $\approx$20 arcsec ($\approx$3.5\,kpc) in VLA data at 5 and 1.5\,GHz \citep{Miyaji1992ApJ...385..137M}. In particular component `B', which lies 2.14 arcsec ($\sim$ 410\,pc) almost directly north of the core before the jet is deflected eastwards, is detected in our tapered radio images. It has a peak flux of 0.61$\pm$0.04\,mJy which is in good agreement with the value 0.64$\pm$0.1\,mJy obtained by \citet[][]{Miyaji1992ApJ...385..137M}. They estimated a size of this component of 1.0$\times$0.6 arcsec$^{2}$, which may explain the detection only in the tapered images with a 140\,mas resolution. 

NGC~4036 is a LINER-type nucleus with an extended east-west jet over 5 arcsec (500\,pc) at 1.5\,GHz \citep{BaldiLeMMINGs2}. The core is fainter than the western component in the 1.5\,GHz image at 0.90$\pm$0.08~mJy and 2.4$\pm$0.1~mJy, respectively. In our 5\,GHz data, the core is coincident with the \textit{Gaia} optical position and in good agreement with the VLBI positions presented in \citet[][]{Cheng2025}. We record a similar flux density to the 1.5\,GHz dataset (0.76$\pm$0.05~mJy) indicating a flat radio spectrum, whereas the western jet component has more than halved in flux density to 0.68$\pm$0.05~mJy compared to the 1.5\,GHz data. Despite having a higher flux density than the `core', the western component is clearly part of the jet and has an angular source size of 69$\times$31 mas$^2$. The total extent in our 5\,GHz data is 2.07 arcsec = 207\,pc. At the resolution and sensitivity of these observations, individual components outside of the central cutout regions are rare, but they are detectable if the emission from large-scale jets is in a compact region that the \textit{e}-MERLIN array is sensitive too, i.e. deconvolved sizes of order the \textit{e}-MERLIN beam.

\begin{figure}
\centering
   \begin{subfigure}[b]{\columnwidth}
   \includegraphics[width=0.95\columnwidth]{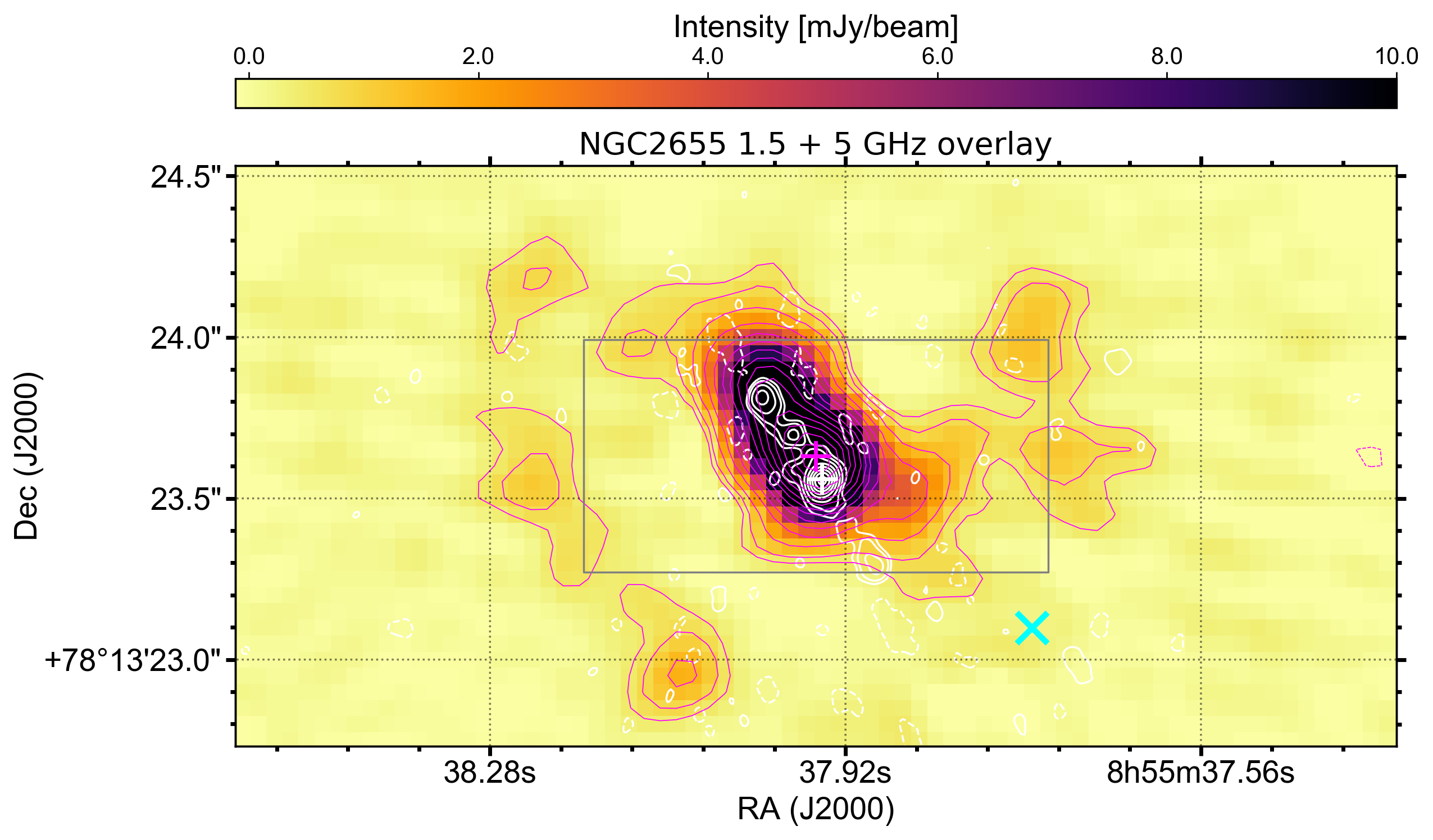}
   \caption{}
   \label{fig:NGC2655Overlaya} 
\end{subfigure}

\begin{subfigure}[b]{\columnwidth}
   \includegraphics[width=0.95\columnwidth]{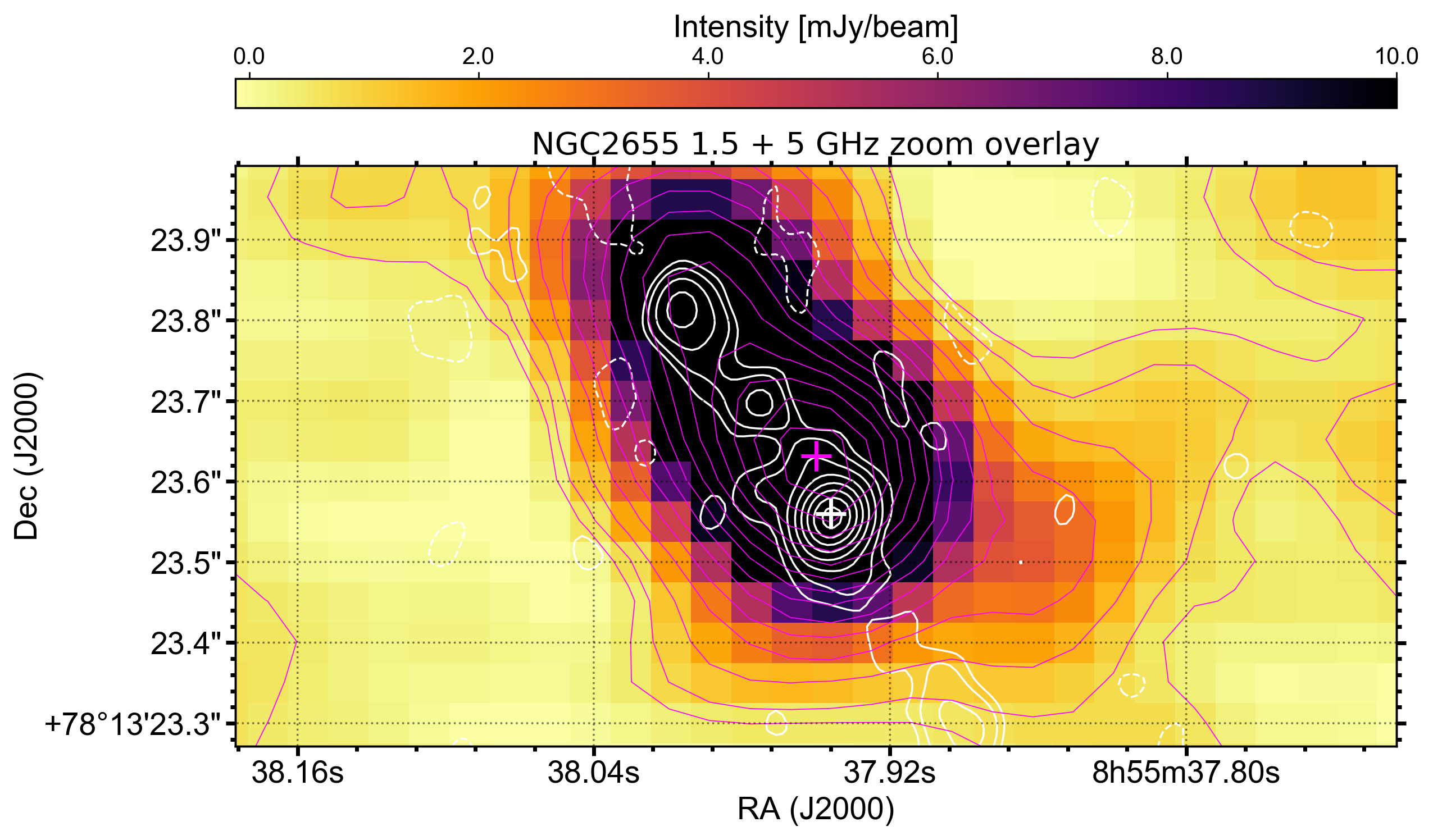}
   \caption{}
   \label{fig:NGC2655Overlayb}
\end{subfigure}
\caption{Similar to Fig.~\ref{fig:NGC6946} and~\ref{fig:NGC507}, but showing the overlay of the 5\,GHz images in this work on top of the 1.5\,GHz images from \citet{BaldiLeMMINGs} for the `jetted' galaxy NGC~2655. The rectangular square in \textit{panel a:} shows the zoomed-in region shown in \textit{panel b:}. The background 1.5\,GHz full-resolution image has magenta contours of $-$3, 3, 5, 9, 16, 25, 36, 49, 64, 81, 100, 121, 144, 169, 196$\sigma$ with 1$\sigma$ equal to 120~$\upmu$Jy beam$^{-1}$. The white contours represent the 5\,GHz data presented in this work at $-$3, 3, 5, 9, 16, 25, 36, 49, 64$\sigma$ with 1$\sigma$ equal to  165~$\upmu$Jy beam$^{-1}$. Negative contours are dashed. The cyan `$\times$' marks the Simbad position for this galaxy, the magenta `+' shows the 1.5\,GHz position and the white `+' denotes the 5\,GHz position. } 
    \label{fig:NGC2655Overlay}

\end{figure}

There are four sources with triple structures: NGC~2655, NGC3504, the aforementioned NGC~4151 and NGC~4589. We discuss NGC~3504 in the context of the `jetted' \ion{H}{ii} galaxies in Section~\ref{sec:jettedh2}. We observe similar radio structures in NGC~2655 and NGC~4589, namely a bright core with multiple radio components either side of the core. At lower-resolution, the radio images show a triple structure centred on the core component. Both are in ETGs and are LINER-type nuclei residing at similar distances (24.4 and 30\,Mpc), resulting in similar angular scales in both sources. NGC~2655 has a complicated `S-shaped' structure at 1.5\,GHz and we only see the central linear component in this 5\,GHz dataset (see example overlay in Fig.~\ref{fig:NGC2655Overlay}, whereas NGC~4589 appears as a single core with a one-sided extension in the 1.5\,GHz data. Interestingly, NGC~2655 was not detected in the VLBA sample of \citet[][]{Nagar2002} to a limit of 0.76~mJy beam$^{-1}$ at $\sim$5\,mas resolution, but it is detected in more recent VLBA observations at 5\,GHz at 0.80$\pm$0.08~mJy beam$^{-1}$ \citep{Cheng2025}. We detect its core with peak intensity 11.11$\pm$0.61\,mJy beam$^{-1}$ indicating significant flux is resolved out between these two resolutions. On the other hand, NGC~4589 is detected in the VLBI study of \citet[][]{nagar05}, but with a much reduced peak intensity: 6.0\,mJy beam$^{-1}$ compared to 17.61$\pm$0.88\,mJy beam$^{-1}$ in our \textit{e}-MERLIN data.

Another aspect which is common between these two triple sources NGC~4589 and NGC~2655 is their high column densities derived from X-ray observations. 
NGC~4589 is listed as a potential Compton Thick AGN candidate \citep[e.g., the nuclear absorption $n_{\rm H}>$1.5$\times$10$^{24}$~cm$^{-2}$, see][]{Gonzalez2015} and NGC~2655 has an $n_{\rm H}\sim$4$\times$10$^{23}$ cm$^{-2}$ \citep{Zhao2021A&A...650A..57Z} from high-quality \textit{NuSTAR} data. Conversely, the vast majority ($>$70 per cent\footnote{The X-ray LeMMINGs campaign  includes a declination-limited sub-sample of the full LeMMINGs survey, so only a lower-limit is given.}) of the LeMMINGs sample have $n_{\rm H}\ll$10$^{22}$~cm$^{-2}$ \citep{X-rayLeMMINGs}, though finding bona-fide AGN is difficult and future \textit{NuSTAR} data is required to confirm highly obscured nuclei. Other LeMMINGs sources with resolved radio structures and high X-ray absorption values are: NGC~2273 \citep{Brightman2017} which shows a westward protrusion in its full-resolution image; NGC~2639 which shows a bright (21\,mJy) core component with a one-sided jet \citep{Gonzalez2015}. NGC~2273, NGC~2639 and NGC~2655 all show very complex morphologies in the 1.5\,GHz sample, with the 5\,GHz data resolving out diffuse larger-scale structures. In addition, NGC~4102 is also heavily obscured \citep{Ricci2015, Marchesi2018} and has a significantly offset 5\,GHz position to the 1.5\,GHz structure, potentially pointing to different emission processes responsible at different scales. High-resolution radio interferometric techniques represent a suitable tool to enable an unobscured view of the active SMBHs in some of these obscured LLAGN.

\vspace{-1em}
\subsubsection{Star formation and `jetted' \ion{H}{II} galaxies}
\label{sec:jettedh2}

In the 1.5\,GHz survey, seven \ion{H}{II} galaxies were labelled as `jetted', and showed characteristics similar to other LLAGN in the multi-wavelength diagnostic plots \citep{BaldiLeMMINGs,BaldiLeMMINGs2,BaldiLeMMINGs3}: NGC~972, NGC~3665, UGC~3828,
NGC~7798, UGC~4028, NGC~2782, and NGC~3504. Out of these seven, three (NGC~2782, NGC~3504 and NGC~3665) are detected in the 5\,GHz data, but they show a variety of morphologies. As noted in Section~\ref{sec:compstudies}, all three of these detected sources were already detected with VLBA/VLBI  \citep{NGC37182007A&A...464..553K,DellerMiddelberg,Liuzzo2009A&A...505..509L} and are possibly LLAGN. However, a non-AGN origin cannot be completely ruled out because their brightness temperatures are up to 10$^6$K. In particular, \citet{Jogee1998} argued that the alignment between the radio and the optical emission lines in NGC~2782 is consistent with a starburst-driven outflow. Whereas the triple structure of NGC~3504 at 5\,GHz and of NGC~3665 at 1.5\,GHz are more suggestive of a jetted system launched from an active SMBH. Thus the detected `jetted' \ion{H}{II} galaxies at 5\,GHz most likely harbour an active SMBH. A forthcoming study with optical, X-ray and emission line data will provide a more conclusive result.

Of the four jetted \ion{H}{II} galaxies undetected in our 5\,GHz data, all bar NGC~7798 had detections below a 5$\sigma$ confidence threshold in the 1.5\,GHz data. Different scenarios are possible: their 1.5-GHz detections were spurious or they have been resolved out by the \textit{uv}-tapering properties of the higher resolution data. Deeper radio follow-up observations are needed to explore their nature.

The detected \ion{H}{II} galaxies with a compact core morphology (`A') could in principle host SF. A SF-driven outflow could explain the ring-like emission in the full-resolution image of NGC~2782. Thermal emission from ultra-compact \ion{H}{II} regions could contribute more at this frequency than at 1.5\,GHz, although the sensitivity level is too poor to detect \ion{H}{II} regions even in the nearest galaxies. For example, the brightest \ion{H}{II} region (source name 42.20+59.1) in the nearby starburst galaxy M82 (NGC3034) \citep{muxlow94,fenech08} is not detected in the widefield images in this work. Thus, in the same vein as the 1.5\,GHz study, we calculate the range of SN rates for these galaxies, assuming that the emission is non-thermal, by using the formula in \citet{condon92}, modified to 5\,GHz: $L_{\rm radio}/(5\times10^{38}$~erg~s$^{-1}) = 6.3\times(\nu_{\rm SN}/\rm yr^{-1})$. Excluding the radio-loud source NGC~3665 and the high brightness temperature obtained from VLBI for NGC~2146, the range of the SN rates for the remaining seven `A' morphology \ion{H}{II} galaxies is 3.2 $\times10^{-5}$ -- 2.0 $\times10^{-3}$ yr$^{-1}$ in a region $\sim$15\,pc around the core. As noted in \citet{BaldiLeMMINGs}, these SN rates are similar to those in M82 for the entire galaxy. Therefore, in our 5-GHz survey, there is no clear evidence of an extreme SN factory based on the radio properties, but some nuclei could be dominated by free-free emission from intense SF \citep{kennicutt12}.

\vspace{-1em}
\subsection{Are there missing LLAGN at 5~GHz?}
\label{sec:detectionLLAGNdiscussion}

It is evident that the \textit{e}-MERLIN data presented in this work preferentially detect compact ($\approx$50\,mas scale) radio emission, although our median sensitivity (66~$\upmu$Jy beam$^{-1}$) is a factor three better than previous radio surveys of the Palomar galaxies  \citep[e.g.,][]{nagar05}. In this section we consider whether we have missed any AGN-dominated sources by not using `deeper' observations.

First, the spectral index may have a significant impact on whether sources are detectable in our 5\,GHz data. As noted previously, LLAGN generally have flat ($\alpha \sim$0) or steep ($\alpha\sim -$0.7) radio spectral indices in the 1$-$10 GHz regime \citep[e.g.,][]{falcke00,Nagar2002,Chiaraluce2019MNRAS.485.3185C}. The VLA FRAMEx project studied the radio spectra of nearby active galaxies \citep{Dorland}, and reported a mean spectral index of $-$0.69 for the twelve detected sources out of their volume-limited sample of twenty five sources \citep{Sargent2025}, many of which have significant overlap with the Palomar sample. Assuming that the reported mean $\alpha$ is characteristic of our LeMMINGs sources (although see Section~\ref{sec:specindex} for a preliminary spectral index analysis which shows the radio cores are consistent with $\alpha$ = 0), then on average a source would need to be $\geq$0.75~mJy beam$^{-1}$ in the 1.5\,GHz data to be detected in our 5\,GHz images. Only 53 sources were detected above 0.75~mJy beam$^{-1}$ in the 1.5\,GHz data \citep{BaldiLeMMINGs,BaldiLeMMINGs2}. Furthermore, if a 5$\sigma$ threshold had been used in the 1.5\,GHz dataset, then only 75 sources would have been detected. Therefore our detection of 68 sources in the 5\,GHz sample is in line with expectations when compared to the 1.5\,GHz data. To improve the sensitivity to a point where all sources detected at 3$\sigma$ in the 1.5\,GHz sample could have been detected in the 5\,GHz sample assuming a radio spectral index of $\alpha =-$0.69, a moderate improvement in sensitivity to 33~$\upmu$Jy beam$^{-1}$ (corresponding to four times longer observations) would be needed. Equivalently, adding the Lovell telescope to these observations would have the same result, but as shown in Figure~\ref{fig:NGC4151comp}, the improved \textit{uv}-coverage of longer observations provides better imaging fidelity compared to deeper `snap-shot' images.

Source variability could also lead to a non-detection in our survey. Seyferts are known to vary over the course of years to decades \citep[e.g., see][]{Mundell2009,Panessa2019,Kharb2026ApJ...997..283K}. 
For example, NGC~3718 has been studied with MERLIN and \textit{e}-MERLIN previously, with total flux densities of 6.1$\pm$0.1~mJy at 5\,GHz \citep[data obtained 2002/2003,][]{NGC37182007A&A...464..553K}, 8.85$\pm$0.07~mJy \citep[August 2011,][]{NGC37182015A&A...580A..11M} and 10.48$\pm$0.53~mJy (LeMMINGs 5\,GHz, August 2019), indicating its long-term flux increase of up to 40 per cent.
The aforementioned NGC~4151 has also been shown to have varied by up to 50 per cent over the last 20 years, but the most likely scenario is the continued interaction between the jet and extended emission line region in this object, as opposed to AGN variability \citep{Williams4151_2}. Variable cores have been also detected in LINERs \citep{2002A&A...385..425F}: we detect NGC~1961 and NGC~7217 at flux densities of 0.86$\pm$0.073~mJy and 0.52$\pm$0.04~mJy, respectively. Both were detected in 5\,GHz MERLIN observations in 2002/2003 at flux densities of 1.2$\pm$0.3~mJy and 5.0$\pm$0.5~mJy \citep{NGC37182015A&A...580A..11M}, respectively, although the origin of this large intensity drop, corresponding to an 8.9$\sigma$ decrease over $\sim$16 years, is unclear. When considering the 1.5\,GHz \textit{e}-MERLIN data, variability may also lead to a non-detection between the two epochs, as the 1.5\,GHz data were observed between 2014--2019, whereas the 5\,GHz data were observed in 2018--2020.

Finally, some sources may simply be too faint to be detectable without significantly more sensitive observations. For example, the Seyfert galaxy NGC~4051 has a triple structure and is clearly detected in our 1.5\,GHz images \citep{BaldiLeMMINGs} with a core peak intensity of 0.38\,mJy beam$^{-1}$. At VLBI resolutions it is also detected, with a core flux density of 0.2\,mJy at 5\,GHz \citep{giroletti09}, below the 5\,GHz \textit{e}-MERLIN 5$\sigma$ detection threshold here. In this case, the aforementioned factor four increase in observing time would likely have led to the detection of this source. Sources with relic AGN activity or low-level emission from SF could similarly be affected as the source becomes too faint and too diffuse to detect at 5\,GHz. For example, the LeMMINGs `deep' sources NGC~6217 and IC~10 revealed radio emission on several arcsecond scales, but both required high ($\leq$10~$\upmu$Jy beam$^{-1}$) sensitivities and excellent \textit{uv}-coverage to detect the resolved features in these galaxy nuclei \citep{Westcott2017,Williams6217}. 
Clearly, high-sensitivity images with more complete $uv$-coverage are required to detect the faintest radio emission at the galaxy centre to unveil a weakly accreting SMBH.

\vspace{-1em}
\section{Conclusions}
\label{sec:conclusions}
We present the 5\,GHz data release of the Legacy \textit{e}-MERLIN Multi-band Imaging of Nearby Galaxies survey (LeMMINGs): a statistically complete census of nuclear activity of 280 nearby galaxies brighter than $B_{\rm T} <$ 12.5 mag, selected from the original Palomar sample \citep{Filippenko85,Ho95,Ho97a,Ho97b,Ho97c,Ho03,ho09}, with a declination of $\delta >$20$^{\circ}$ for radio imaging fidelity. The survey consists of radio continuum observations of all types of optical nuclei and galaxy types. This work provides a crucial second frequency band to the 1.5\,GHz data already published \citep{BaldiLeMMINGs,BaldiLeMMINGs2,BaldiLeMMINGs3} to help diagnose the emission mechanisms responsible for the nuclear properties of these galaxies. We focus on the new 5\,GHz radio data, providing 50\,mas resolution images of the entire sample down to a median sensitivity of 66~$\upmu$Jy beam$^{-1}$ and a 5$\sigma$ detection threshold of 0.33~mJy beam$^{-1}$. This sensitivity represents a factor of three improvement when compared to previous investigations of the Palomar sample \citep{filho00,Nagar2002,nagar05}. We also provide tapered images at a resolution of 0.14 arcseconds, matched to the 1.5\,GHz LeMMINGs campaign.

In total, we detect nuclear radio emission within 1.5-arcsec of either the \textit{Gaia} or Simbad optical positions in 68/280 galaxies, consistent with previous radio studies of the Palomar sample \citep{nagar05,HoReview}. We use a search radius of 1.5 arcsec around the \textit{Gaia} position (224/280 galaxies), if it agreed with the archival position of the galaxy nucleus, or the Simbad position if they were discrepant. In most cases, we detect radio emission within 1.5 arcsec of the \textit{Gaia} position, but in eight cases the Simbad position was used to locate the nuclear region. The comparison to the \textit{Gaia} positions shows that the \textit{Gaia} optical co-ordinates can be useful for searching for nuclear radio emission. Care must be taken when using the \textit{Gaia} positions for edge-on galaxies or lenticular galaxies, where the positions appear to be more offset than in other galaxy types, possibly due to misidentification of the bright nuclear region. Considering the 1.5\,GHz and 5\,GHz samples together as one, we find a total of 79 radio-detected nuclei with signatures of radio AGN activity within the sample of 280 galaxies. This value implies that approximately 30 \textit{per cent} of galaxies in the local Universe host a SMBH capable of detectable radio emission at $\sim$100s $\upmu$Jy levels.

One galaxy (NGC~5194) was only detected in tapered images. Two galaxies (NGC~3642 and NGC~4138) were not detected in the 1.5\,GHz LeMMINGs sample, which may be due to nuclear variability or possibly inverted radio spectra. In most cases (58/65), the 5\,GHz nuclear radio emission agrees to within 0.2 arcsec with the 1.5\,GHz LeMMINGs positions, but in a handful of cases a shift greater than 0.2 arcsec is observed, mostly due to unresolved radio components in the 1.5\,GHz images or misidentification of the radio nucleus from the lower-resolution 1.5\,GHz data. 

The median luminosity of the detected sources in the sample is 3.6$\times$10$^{36}$~erg s$^{-1}$ (7.2$\times$10$^{19}$~W Hz$^{-1}$, the deepest sub-arcsec 5-GHz survey of the Palomar galaxies), though the censored mean luminosities for all classes are lower than this value, due to the large number of non-detections in the sample. The galaxies most commonly detected in the radio are the `active' galaxies, i.e., the Seyferts and LINERs (53/112, 47 per cent). The radio detected galaxies mostly reside in early-type galaxies (40 per cent), with LINERs and Seyferts dominating both the detection fraction and the high‑luminosity tail. Many of these sources appear to be genuine LLAGN, as shown by their high brightness temperatures ($T_{\rm B} \geq$10$^{6}$K) in VLBI observations \citep{nagar05,NGC37182007A&A...464..553K,Liuzzo2009A&A...505..509L,Panessa2013,DellerMiddelberg,Cheng2025}. In contrast, the `inactive' galaxies (absorption line galaxies (ALGs) and \ion{H}{II} galaxies) have much lower detection rates (14/168, 8 per cent) and generally at much lower luminosities. Whereas the detected ALGs in ellipticals have similar properties to those of LINERs in support of our previous LeMMINGs studies of ALGs \citep{BaldiLeMMINGs2,X-rayLeMMINGs}, the \ion{H}{II} galaxies are a mixed bag with nuclear radio emission that could be attributed to SF, SN remnants, or LLAGN \citep[e.g.,][]{2000A&A...358...95T,Cheng2025}. We estimate that 3$-$4 per cent of \ion{H}{II} galaxies may host an active SMBH.

We classify the sources into different radio morphological types, with most (52/68) sources showing a compact core component (radio morphology type `A'), and the rest (15/68) showing evidence for extended emission other than a radio core (radio morphology type `B', `C' or `E'), possibly related to a jet. We detect radio structures between 5 and 380 pc in size in these `jetted' sources. The low percentage of `jetted' sources (22 per cent) in the 5\,GHz data is in contrast with the 38 per cent of sources with `jetted' morphologies at 1.5\,GHz. We find that the `jetted' radio structures are observed less frequently in the 5\,GHz LeMMINGs data than in the 1.5\,GHz data for a number of reasons: i) the higher resolution of the 5\,GHz data resolves out the jets in some cases; ii) the `shallow' observations are not sensitive enough to detect the fainter jets arising from optically-thin synchrotron emission; and/or, iii) low dynamic range prevents from
imaging the jets in bright AGN such as in NGC~315 and NGC~1275. In conclusion, with a heterogeneous interferometer such as \textit{e}-MERLIN, we estimate that a factor of four increase in observing time above the observations in this work (equivalent to sensitivities $\leq$33~$\upmu$Jy beam$^{-1}$) with excellent \textit{uv}-coverage is desired to detect jets in nearby galaxies. This would also ensure detections of known LLAGN and allow the search for low-level radio emission from non-AGN driven processes. 

Collectively, our results support the view that LLAGN represents the most common mode of SMBH activity in nearby galaxies at sub-arcsecond scales. We find that the physical drivers of radio emission diverge based on host morphology: in early-type galaxies, the high detection rates, elevated luminosities in LINERs and Seyferts, and the predominance of compact  ($\lesssim$10\,pc) cores at 5\,GHz identify LLAGN as the primary central engines. In contrast, the comparatively low detection rates in \ion{H}{II} and ALG systems suggest that non-AGN processes dominate the nuclear radio emission in many late-type galaxies, with only a small subset of \ion{H}{II} galaxies harbouring bona fide LLAGN. These findings highlight the necessity of high-resolution sensitive radio observations to disentangle genuine nuclear activity from SF processes at pc scales and to obtain an unbiased census of low-luminosity SMBH accretion in the local Universe.

\vspace{-1em}
\section*{Acknowledgements}

We thank the anonymous referee who provided constructive comments that have improved this manuscript.

DWB would like to thank Sam Connolly, Dimitrios Emmanolopoulos, Alasdair Thomson, Bob Watson, Sara Motta, Lauren Rhodes for useful discussions. 
We acknowledge Jodrell Bank Centre for Astrophysics, which is funded by the STFC. \textit{e}-MERLIN and formerly, MERLIN, is a National Facility operated by the University of Manchester at Jodrell Bank Observatory on behalf of STFC. This work has made use of data from the European Space Agency (ESA) mission {\it Gaia} (\url{https://www.cosmos.esa.int/gaia}), processed by the {\it Gaia} Data Processing and Analysis Consortium (DPAC, \url{https://www.cosmos.esa.int/web/gaia/dpac/consortium}). Funding for the DPAC has been provided by national institutions, in particular the institutions participating in the {\it Gaia} Multilateral Agreement. This research has made use of the NASA/IPAC Extragalactic Database (NED), which is operated by the Jet Propulsion Laboratory, California Institute of Technology, under contract with the National Aeronautics and Space Administration. This research has made use of the SIMBAD database, CDS, Strasbourg Astronomical Observatory, France \citep{2000A&AS..143....9W}. This research made use of APLpy, an open-source plotting package for Python \citep[][]{APLpy}.

RDB acknowledges financial support from INAF mini-grant \textit{\lq\lq FR0 radio galaxies\rq\rq} (Bando Ricerca Fondamentale INAF 2022). IM acknowledges the Development in Africa with Radio Astronomy (Phase 3) for funding his postdoctoral fellowship through the UK’s Science and Technologies Facilities Council (STFC) grant ST/Y006100/1. IMcH thanks the Royal Society for the award of a Royal Society Leverhulme Trust Senior Research Fellowship. 
MP acknowledges Royal Society-SERB Newton International Fellowship support funded jointly by the Royal Society, UK and the Science and Engineering Board of India (SERB) through Newton-Bhabha Fund. RDB and IMcH also acknowledge the support of STFC under grant [ST/M001326/1]. 
SM gratefully acknowledges the support from the Chandra grant GO7-18080X  issued by the Chandra X-ray Center, which is operated by the Smithsonian Astrophysical Observatory for and on behalf of the National Aeronautics Space Administration under contract NAS8-03060.
Co-funded by the European Union (MSCA Doctoral Network EDUCADO, GA 101119830 and Widening Participation, ExGal-Twin, GA 101158446). JHK acknowledges grant PID2022-136505NB-I00 funded by MCIN/AEI/10.13039/501100011033 and EU, ERDF. 
AA and MAPT acknowledge support from the Spanish MCIU through grant PGC2018-098915-B-C21. JM acknowledges support from the grant RTI2018-096228-B-C31 (MICIU/FEDER, EU). AA, MPT and JM acknowledge financial support through the Severo Ochoa grant CEX2021-001131-S and the Spanish National grant PID2023-147883NB-C21, funded by MCIU/AEI/ 10.13039/501100011033, as well as support through ERDF/EU. MPS acknowledges financial support from MCIU under the Severo Ochoa  grant PRE2021-100265. MPS acknowledges financial support from MCIU under the Severo Ochoa grant PRE2021-100265 and the I+D+i project PID2022-140871NB-C21, financed by MICIU/AEI/10.13039/501100011033/ and ``FEDER/UE''.
PK acknowledges the support of the Department of Atomic Energy, Government of India, under the project 12-R\&D-TFR-5.02-0700. 
AB is grateful to the Royal Society, U.K., and acknowledges SERB (SB/SRS/2022-23/124/PS) for financial support.
XC was supported by the Brain Pool Program through the National Research Foundation of Korea (NRF) funded by the Ministry of Science and ICT (RS-2024-00407499). 
PK acknowledges the support of the Department of Atomic Energy, Government of India, under the project 12- R$\&$D-TFR-5.02-0700. 
FP acknowledges financial support from the Bando Ricerca Fondamentale INAF and "Programma di Ricerca Fondamentale INAF 2023 and 2024".


\vspace{-1em}
\section*{Data Availability}

The raw and calibrated data included in this manuscript is part of the \emerlin{} legacy programmes, and all the 1.5 and 5\,GHz \textsc{CASA} calibrated datasets for LeMMINGs can be found here: \url{https://www.e-merlin.ac.uk/distribute/LEGACY/LE1011/LE1011.html}. We have placed all the fits images of the cut out regions around the sources on the \textit{e}-MERLIN webpage here: \url{https://www.e-merlin.ac.uk/legacy-lemmings.html}. Furthermore, a singular download of all the images in this work and the singularity \textsc{wsclean} container can be found on Zenodo: \url{10.5281/zenodo.17940946}. The tables provided in this work are included in a machine readable format in the online supplementary material and the above Zenodo link.



\bibliographystyle{mnras}

\bibliography{thesisbib2.bib}



\vspace{3mm}
\noindent
{\small 
$^{1}$ Jodrell Bank Centre for Astrophysics, School of Physics and Astronomy, The University of Manchester, Manchester, M13 9PL, UK\\
$^{2}$ INAF - Istituto di Radioastronomia, Via P. Gobetti 101, I-40129 Bologna, Italy\\
$^{3}$ School of Physics and Astronomy, University of Southampton, Southampton, SO17 1BJ, UK\\
$^{4}$ Department of Physical Sciences, Embry-Riddle Aeronautical University, Daytona Beach, FL 32114, USA\\
$^{5}$ Department of Astronomy and Space Science, Technical University of Kenya, P.O Box 52428 - 00200, Nairobi, Kenya\\
$^{6}$ Jeremiah Horrocks Institute, School of Engineering and Computing, University of Lancashire, Preston PR1 2HE, UK\\
$^{7}$ Cahill Center for Astrophysics, California Institute of Technology, 1216 East California Boulevard, Pasadena, CA 91125, USA \\
$^{8}$ Centre for Astrophysics Research, University of Hertfordshire, College Lane, Hatfield, AL10 9AB, UK\\
$^{9}$  SKAO, Jodrell Bank, Lower Withington, Macclesfield, SK11 9FT, UK \\
$^{10}$ Instituto de Astrof\'{i}sica de Canarias, V\'{i}a L\'{a}ctea S/N, E-38205 La Laguna, Spain\\
$^{11}$ Departamento de Astrof\'{i}sica, Universidad de La Laguna, E-38206 La Laguna, Spain\\
$^{12}$ Astronomy department, The Ohio State University, Columbus, OH, 43210, USA\\
$^{13}$ Center for Astronomy and Astro-particle Physics, The Ohio State University, Columbus, OH 43210, USA.  \\
$^{14}$ Eureka Scientific, 2452 DELMER ST STE 100, Oakland, CA, 94602, USA \\
$^{15}$ Instituto de Astrof\'isica de Andaluc\'ia (IAA-CSIC), Glorieta de la Astronom\'ia s/n, 18008 Granada, Spain\\
$^{16}$ Department of Physics, Indian Institute of Technology, Hyderabad 502285, India\\
$^{17}$ Xinjiang Astronomical Observatory, Chinese Academy of Sciences, Urumqi 830011, China\\
$^{18}$ Netherlands Institute for Radio Astronomy (ASTRON), Oude Hoogeveensedijk 4, 7991 PD Dwingeloo, The Netherlands\\
$^{19}$ Indian Institute of Science Education and Research (IISER), Mohali, Punjab, 140306, India\\
$^{20}$ Indian Institute of Astrophysics, Koramangala II Block, Bangalore-560034, India\\
$^{21}$ Korea Astronomy and Space Science Institute, 776 Daedeok-daero, Yuseong-gu, Daejeon 34055, Korea\\
$^{22}$ Astrophysics Group, Cavendish Laboratory, J. J. Thomson Avenue, Cambridge CB3 0US, UK\\
$^{23}$ United Kingdom SKA Regional Centre (UKSRC), UK\\
$^{24}$ National Centre for Radio Astrophysics (NCRA) - Tata Institute of Fundamental Research (TIFR),  Ganeshkhind, Pune 411007, Maharashtra, India\\
$^{25}$ Department of Astrophysics/IMAPP, Radboud University, P.O. Box 9010, 6500GL Nijmegen, The Netherlands\\
$^{26}$ INAF - Istituto di Astrofisica e Planetologia Spaziali, via Fosso del Cavaliere 100, I-00133 Roma, Italy \\
$^{27}$ Institute of Astronomy and Astrophysics, Academia Sinica, 11F of AS/NTU Astronomy-Mathematics Building, No.1, Sec. 4, Roosevelt Rd, Taipei 106319, Taiwan, R.O.C\\
$^{28}$ Fakultat f\"ur Physik, Universit\"at Bielefeld, Postfach 100131, D-33501 Bielefeld, Germany\\
$^{29}$ Assam Don Bosco University, Guwahati 781017, Assam, India\\
$^{30}$ Center for Astro, Particle and Planetary Physics (CAP$^3$), New York University Abu Dhabi, PO Box 129188, Abu Dhabi, UAE\\
$^{31}$ Department of Astronomy, Yale University, PO Box 208101, New Haven, CT 06520-8101, USA\\
$^{32}$ School of Physics and Astronomy, University of Birmingham, Edgbaston, Birmingham B15 2TT, UK\\

}










\bsp	
\label{lastpage}
\end{document}